\documentclass[11pt,nodots,reqno,eqnsec]{style/orca}

\pdfoutput=1

\usepackage{bookmark}
\usepackage{caption}
\usepackage{graphicx}
\usepackage{subcaption}
\usepackage{ytableau}
\usetikzlibrary{arrows.meta, decorations.markings, calc, positioning, decorations.pathmorphing, decorations.shapes, shapes.geometric, calligraphy, patterns}
\tikzset{->-/.style={decoration={
  markings,
  mark=at position .5 with {\arrow{>}}},postaction={decorate}}}
\tikzset{-<-/.style={decoration={
  markings,
  mark=at position .5 with {\arrow{<}}},postaction={decorate}}}
\tikzset{-->-/.style={decoration={
  markings,
  mark=at position .75 with {\arrow[scale=0.75]{>}}},postaction={decorate}}}
\tikzset{--<-/.style={decoration={
  markings,
  mark=at position .35 with {\arrow[scale=0.75]{<}}},postaction={decorate}}}
\tikzset{->--/.style={decoration={
  markings,
  mark=at position .5 with {\arrow[scale=0.75]{>}}},postaction={decorate}}}
\tikzset{-<--/.style={decoration={
  markings,
  mark=at position .5 with {\arrow[scale=0.75]{<}}},postaction={decorate}}}
\ytableausetup{mathmode, boxsize=18pt, aligntableaux=center, centerboxes}

\setlist{itemsep=3pt plus 1pt minus 1pt,topsep=0pt}
\DeclareSymbolFont{Letters}{OML}{cmm}{m}{it}
\DeclareMathSymbol{\psi}{\mathalpha}{Letters}{"20}
\newtheoremstyle{dotless}{}{}{\itshape}{}{\bfseries}{}{ }{}
\theoremstyle{dotless}

\newcommand{\Le}{\textnormal{\reflectbox{L}}}
\newcommand{\BLe}{\textnormal{\textbf{\reflectbox{L}}}}

\newcounter{tpointn}[section]
\renewcommand{\thetpointn}{%
  \ifnum\value{section}=0 %
    \else\thesection.\arabic{tpointn}\fi
  }
\renewcommand{\tpointn}[2][]{\vspace{2mm}\par%
  \ifx&#1&\refstepcounter{tpointn}\fi%
  \noindent{\bf #2 \ifx&#1&\thetpointn\fi\if@dots.\fi\, ---}}

\newcommand{\lebox}[1]{%
  \,\begin{ytableau}#1\end{ytableau}\,
}

\newcommand{\leplus}{\,+\,}

\newcommand{\leconfig}{%
  \ytableausetup{baseline}
  \begin{ytableau}
    \none & \none & + \\
    \none & \none & \none[\vdots] \\
    + & \none[\cdots] & 0
  \end{ytableau}
}

\newcommand{\smallelbow}{%
  \begin{tikzpicture}[scale=0.13]
    \draw (0, 1) arc[start angle=-90, end angle=0, x radius=1, y radius=-1];
    \draw (2, 1) arc[start angle=90, end angle=180, x radius=1, y radius=-1];
  \end{tikzpicture}
}

\newcommand{\smallcross}{%
  \begin{tikzpicture}[scale=0.13]
    \draw (0, 1) to (2,1);
    \draw (1, 0) to (1,2);
  \end{tikzpicture}
}

\newcommand{\elbow}[2][]{%
  \begin{tikzpicture}[scale=0.32]
    \draw[#1] (0, 1) arc[start angle=-90, end angle=0, x radius=1, y radius=-1];
    \draw[#2] (2, 1) arc[start angle=90, end angle=180, x radius=1, y radius=-1];
  \end{tikzpicture}
}

\newcommand{\cross}[2][]{%
  \begin{tikzpicture}[scale=0.32]
    \draw[#1] (0, 1) to (2,1);
    \draw[#2] (1, 0) to (1,2);
  \end{tikzpicture}
}

\newcommand{\exle}{%
  \begin{tikzpicture}[baseline=(current bounding box.center)]
    \useasboundingbox (0,0) rectangle (5*18pt+5*0.04em, 3*18pt+3*0.04em);
    \draw[dotted] (0,0) rectangle (5*18pt+5*0.04em,3*18pt+3*0.04em);
    \node[inner sep=0] at (3*18pt+3*0.04em-9.19pt,2*18pt+2*0.04em-9.19pt) {%
    \begin{ytableau}
      0 & 0 & + & + & 0 \\
      + & + & + & \none & \none \\
      0 & \none & \none & \none & \none 
    \end{ytableau}};
    \draw[decorate,decoration={calligraphic brace,amplitude=6pt,raise=4pt}] (current bounding box.north west) -- (current bounding box.north east) node [midway, above=10pt] {\scriptsize$5=8-3$};
    \draw[decorate,decoration={calligraphic brace,mirror,amplitude=6pt,raise=4pt}] (current bounding box.north west) -- (current bounding box.south west) node [midway, left=10pt] {\scriptsize$3$};
  \end{tikzpicture}
}

\newcommand{\expipe}{%
  \begin{tikzpicture}[baseline=(current bounding box.center)]
    \useasboundingbox (0,0) rectangle (5*18pt+5*0.04em, 3*18pt+3*0.04em);
    \draw[dotted] (0,0) rectangle (5*18pt+5*0.04em,3*18pt+3*0.04em);
    \node[inner sep=0] at (3*18pt+3*0.04em-9.18pt,2*18pt+2*0.04em-9.18pt) {%
    \begin{ytableau}
      \cross{} & \cross{} & \elbow{->--} & \elbow{->--} & \cross{-->-} \\
      \elbow{->--} & \elbow{->--} & \elbow{-->-} & \none & \none \\
      \cross[--<-]{-->-} & \none & \none & \none & \none 
    \end{ytableau}};
    \node[inner sep=0] at (-3pt, 9pt+0.04em) {\tiny$7$};
    \node[inner sep=0] at (-3pt, 3*9pt+2*0.04em) {\tiny$4$};
    \node[inner sep=0] at (-3pt, 5*9pt+3*0.04em) {\tiny$1$};
    \node[inner sep=0] at (9pt+0.04em, 3*18pt+3*0.04em+3pt) {\tiny$8$};
    \node[inner sep=0] at (3*9pt+2*0.04em, 3*18pt+3*0.04em+3pt) {\tiny$6$};
    \node[inner sep=0] at (5*9pt+3*0.04em, 3*18pt+3*0.04em+3pt) {\tiny$5$};
    \node[inner sep=0] at (7*9pt+4*0.04em, 3*18pt+3*0.04em+3pt) {\tiny$3$};
    \node[inner sep=0] at (9*9pt+5*0.04em, 3*18pt+3*0.04em+3pt) {\tiny$2$};
    \node[inner sep=0] at (1*18pt+1*0.04em+3pt, 9pt+0.04em) {\tiny$7$};
    \node[inner sep=0] at (3*18pt+3*0.04em+3pt, 3*9pt+2*0.04em) {\tiny$4$};
    \node[inner sep=0] at (5*18pt+5*0.04em+3pt, 5*9pt+3*0.04em) {\tiny$1$};
    \node[inner sep=0] at (3*9pt+2*0.04em, 1*18pt+1*0.04em-4pt) {\tiny$6$};
    \node[inner sep=0] at (5*9pt+3*0.04em, 1*18pt+1*0.04em-4pt) {\tiny$5$};
    \node[inner sep=0] at (7*9pt+4*0.04em, 2*18pt+2*0.04em-4pt) {\tiny$3$};
    \node[inner sep=0] at (9*9pt+5*0.04em, 2*18pt+2*0.04em-4pt) {\tiny$2$};
    \node[inner sep=0] at (9pt+0.04em, -4pt) {\tiny$8$};
  \end{tikzpicture}
}

\newcommand{\exnotation}{%
  \begin{tikzpicture}[baseline=(current bounding box.center)]
    \node[inner sep=0] at (3*18pt+3*0.04em-9.19pt,2*18pt+2*0.04em-9.19pt) {%
    \begin{ytableau}
      0 & 0 & + & + & 0 \\
      + & + & + & \none & \none \\
      0 & \none & \none & \none & \none  \\
    \end{ytableau}};
    \node[inner sep=0] at (-3pt, 9pt+0.04em) {\tiny$7$};
    \node[inner sep=0] at (-3pt, 3*9pt+2*0.04em) {\tiny$4$};
    \node[inner sep=0] at (-3pt, 5*9pt+3*0.04em) {\tiny$1$};
    \node[inner sep=0] at (9pt+0.04em, 3*18pt+3*0.04em+3pt) {\tiny$8$};
    \node[inner sep=0] at (3*9pt+2*0.04em, 3*18pt+3*0.04em+3pt) {\tiny$6$};
    \node[inner sep=0] at (5*9pt+3*0.04em, 3*18pt+3*0.04em+3pt) {\tiny$5$};
    \node[inner sep=0] at (7*9pt+4*0.04em, 3*18pt+3*0.04em+3pt) {\tiny$3$};
    \node[inner sep=0] at (9*9pt+5*0.04em, 3*18pt+3*0.04em+3pt) {\tiny$2$};
    \node[inner sep=0] at (1*18pt+1*0.04em+3pt, 9pt+0.04em) {\tiny$7$};
    \node[inner sep=0] at (3*18pt+3*0.04em+3pt, 3*9pt+2*0.04em) {\tiny$4$};
    \node[inner sep=0] at (5*18pt+5*0.04em+3pt, 5*9pt+3*0.04em) {\tiny$1$};
    \node[inner sep=0] at (3*9pt+2*0.04em, 1*18pt+1*0.04em-4pt) {\tiny$6$};
    \node[inner sep=0] at (5*9pt+3*0.04em, 1*18pt+1*0.04em-4pt) {\tiny$5$};
    \node[inner sep=0] at (7*9pt+4*0.04em, 2*18pt+2*0.04em-4pt) {\tiny$3$};
    \node[inner sep=0] at (9*9pt+5*0.04em, 2*18pt+2*0.04em-4pt) {\tiny$2$};
    \node[inner sep=0] at (9pt+0.04em, -4pt) {\tiny$8$};
    \draw[->] (3,2.35) to (0.25,2.35);
    \draw[->] (-0.4,1.75) to (-0.4,0.25);
  \end{tikzpicture}
}

\newcommand{\constructD}{%
  \begin{tikzpicture}[main node/.style={inner sep=0,minimum size =.04cm,circle,fill=black!80,draw},node distance = 0.75cm,scale=0.5]
    \begin{scope}
      \draw (0,0) -- ++(0,4) -- ++(6,0) -- ++(0,-1) node[midway,right] {\vdots} -- ++(-1,0) -- ++(0,-1) node[midway,right] {\tiny$b_{j-1}$} -- ++(-2,0) -- ++(0,-1) node[midway,right] {\tiny$b_{j}$} -- ++(-1,0)
      -- ++(0,-1) node[midway,right] {\vdots} -- cycle;
    \node at (2,2.5) {$\hat{D}$};
    \draw[pattern=north east lines, pattern color=gray!50] (3.85,2) rectangle (4.35,4) ;
    \node[below] at (4.1,2) {\tiny $a_n$};
    \end{scope}
    \node at (7,2) {$\longrightarrow$};
    \begin{scope}[shift={(8,0)}]
      \draw (0,0) -- ++(0,4) -- ++(5.5,0) -- ++(0,-1) node[midway,right] {\vdots} -- ++(-1,0) -- ++(0,-1) node[midway,right] {\tiny$b_{j-1}$} -- ++(-1.5,0) -- ++(0,-1) node[midway,right] {\tiny$b_{j}$} -- ++(-1,0)
      -- ++(0,-1) node[midway,right] {\vdots} -- cycle;
      \draw[dashed] (3.85,2) to (3.85,4);
    \end{scope}
    \node at (14.5,2) {$\longrightarrow$};
    \begin{scope}[shift={(15.5,-0.25)}]
      \draw (0,0) -- ++(0,4.5) -- ++(5.5,0) -- ++(0,-1) node[midway,right] {\vdots} -- ++(-1,0) -- ++(0,-1) node[midway,right] {\tiny$b_{j-1}$} -- ++(-0.75,0) -- ++(0,-0.5) node[above=2pt,right] {\tiny$a_n$} -- ++(-0.75,0) -- ++(0,-1) node[midway,right] {\tiny$b_{j}$} -- ++(-1,0)
      -- ++(0,-1) node[midway,right] {\vdots} -- cycle;
      \node at (2,3.5) {$D$};
    \draw[pattern=north east lines, pattern color=gray!50] (0,2) rectangle (3.75,2.5) ;
    \end{scope}
  \end{tikzpicture}
}

\newcommand{\steptwoii}{%
  \begin{tikzpicture}[main node/.style={inner sep=0,minimum size =.04cm,circle,fill=black!80,draw},node distance = 0.75cm,scale=0.7]
    \draw (-0.15,-0.15) -- ++(0,-4*0.85) -- ++(4,0) -- ++(0,0.85) -- ++(3,0) -- ++(0,0.85) -- ++(1,0) -- ++(0,0.85) -- ++(1,0) -- ++(0,0.85) node[midway,right] {\small$b_u$};
    \node at (-2,-2) {$\hat{D}$};
    \node at (0.25, -0.5) {$0$};
    \node at (0.85, -0.5) {$\rule{15pt}{0.5pt}$};
    \node at (1.45, -0.5) {$0$};
    \node at (1.95, -0.5) {$\leplus$};
    \node at (1.95, -4.0) {$L_u$};

    \node at (0.25, -1.35) {$0$};
    \node at (2.35, -1.35) {$\rule{70pt}{0.5pt}$};
    \node at (4.45, -1.35) {$0$};
    \node[draw,circle,inner sep=1pt] at (5.1, -1.35) {$\leplus$};
    \fill[pattern=north east lines, pattern color=gray!50] (3.85,-2.7) rectangle (4.65,0) ;

    \node at (0.25, -2.2) {$0$};
    \node at (3.00, -2.2) {$\rule{90pt}{0.5pt}$};
    \node at (5.75, -2.2) {$0$};
    \node at (6.25, -2.2) {$\leplus$};

    \fill[pattern=north east lines, pattern color=gray!50] (2.2,-3.55) rectangle (2.65,0) ;
    \node[draw,circle,inner sep=1pt] at (3.1, -3.05) {$\leplus$};
  \end{tikzpicture}
}

\newcommand{\leconfiglabels}{%
  \ytableausetup{boxsize=1.75em,baseline}\small
  \begin{ytableau}
    \none & \none & + & \none[\leftarrow] & \none[\text{row } $i$]\\
    \none & \none & \none[\vdots] \\
    + & \none[\cdots] & 0 & \none[\leftarrow] & \none[\text{row } $j$] \\
    \none[\text{col. } $m$] & \none & \none[\text{col. } \ell]
  \end{ytableau}
}

\newcommand{\algexDshapes}{%
  \begin{tikzpicture}[scale=0.85]
    \pgfmathsetmacro{\unit}{0.7};
    \begin{scope}
      \draw (-0.25,0) -- ++(0,4*\unit) -- ++(12*\unit+0.25,0) -- ++(0,-1*\unit) node[midway,right] {\tiny{$1$}} -- ++(-1.05*\unit,0) -- ++(0,-1*\unit) node[midway,right] {\tiny{$i$}} -- ++(-3.35*\unit,0) -- ++(0,-1*\unit) node[midway,right] {\tiny$o$} -- ++(-1.8*\unit,0) -- ++(0,-1*\unit) node[midway,right] {\tiny{$q$}} -- cycle;
      \node at (2, -1.5) {\small{$\hat{D}$}};
      \node at (0, -0.25*\unit) {\tiny$n$};
      \node at (0.75, -0.25*\unit) {\tiny$u$};
      \node at (1.5, -0.25*\unit) {\tiny$t$};
      \node at (2.5, -0.25*\unit) {\tiny$s$};
      \node at (3.25, -0.25*\unit) {\tiny$r$};
      \node at (4.75, 0.75*\unit) {\tiny$p$};
      \node at (6, 1.75*\unit) {\tiny$m$};
      \node at (6.75, 1.75*\unit) {\tiny$\ell$};
      \node at (0.75, 3.5*\unit) {\small$+$};
      \node at (6, 3.5*\unit) {\small$+$};
      \node at (6.75, 2.5*\unit) {\small$+$};
      \node at (2.5, 1.5*\unit) {\small$+$};
      \node at (4.75, 1.5*\unit) {\small$+$};
      \node at (1.5, 0.5*\unit) {\small$+$};
      \node at (2.5, 0.5*\unit) {\small$+$};
      \node at (3.25, 0.5*\unit) {\small$+$};
    \end{scope}
    \begin{scope}[shift={(9.75,-0.5)}]
    \draw (-0.25,0) -- ++(0,5*\unit) -- ++(12*\unit+0.25,0) -- ++(0,-1*\unit) node[midway,right] {\tiny{$1$}} -- ++(-1.05*\unit,0) -- ++(0,-1*\unit) node[midway,right] {\tiny{$i$}} -- ++(-2.15*\unit,0) --++(0,-1*\unit) node[midway, right] {\tiny$m$} -- ++(-1.05*\unit,0) -- ++(0,-1*\unit) node[midway,right] {\tiny$o$} -- ++(-1.8*\unit,0) -- ++(0,-1*\unit) node[midway,right] {\tiny{$q$}} -- cycle;
    \node at (2, -1) {\small{${D}$}};
    \node at (0, -0.25*\unit) {\tiny$n$};
    \node at (0.75, -0.25*\unit) {\tiny$u$};
    \node at (1.5, -0.25*\unit) {\tiny$t$};
    \node at (2.5, -0.25*\unit) {\tiny$s$};
    \node at (3.25, -0.25*\unit) {\tiny$r$};
    \node at (4.75, 0.75*\unit) {\tiny$p$};
    \node at (6.75, 2.75*\unit) {\tiny$\ell$};
    \draw[dotted] (0.625,5*\unit) -- ++(0.25,0) -- ++(0,-2*\unit) -- ++(0.5,0) -- ++(0,-1*\unit) -- ++ (-0.75,0) -- cycle;
    \draw[dotted] (1.375,3*\unit) -- ++(0.25,0) -- ++(0,-2*\unit) -- ++(0.75,0) -- ++(0,-1*\unit) -- ++ (-1,0) -- cycle;
    \draw[dotted] (2.375,3*\unit) -- ++(0.25,0) -- ++(0,-2*\unit) -- ++(0.5,0) -- ++(0,-1*\unit) -- ++ (-0.75,0) -- cycle;
    \draw[dotted] (3.125,2*\unit) -- ++(0.25,0) -- ++(0,-1*\unit) -- ++(0.79,0) -- ++(0,-1*\unit) -- ++ (-1.04,0) -- cycle;
    \draw[dotted] (4.625,3*\unit) -- ++(0.25,0) -- ++(0,-1*\unit) -- ++(0.54,0) -- ++(0,-1*\unit) -- ++ (-0.79,0) -- cycle;
    \draw[dotted] (6.625,5*\unit) -- ++(0.25,0) -- ++(0,-1*\unit) -- ++(0.79,0) -- ++(0,-1*\unit) -- ++ (-1.04,0) -- cycle;
    \begin{scope}[shift={(0,0)}]
      \node at (0.75, 4.5*\unit) {\small$+$};
      \node at (6.2, 4.5*\unit) {\small$+$};
      \node at (6.35, 4.47*\unit) {\tiny$\cdot$};
      \node at (6.5, 4.5*\unit) {\small$+$};
      \node at (6.75, 4.5*\unit) {\small$+$};
      \node at (7.75, 4.5*\unit) {\small$+$};
      \node at (8.0, 4.47*\unit) {\tiny$\cdots$};
      \node at (8.25, 4.5*\unit) {\small$+$};
    \end{scope}
    \begin{scope}[shift={(0,4*\unit)}]
      \node at (0.75, -0.5*\unit) {\small$0$};
      \node at (6.75, -0.5*\unit) {\small$0$};
      \node at (7, -0.5*\unit) {\small$+$};
      \node at (7.25, -0.53*\unit) {\tiny$\cdots$};
      \node at (7.5, -0.5*\unit) {\small$+$};
    \end{scope}
    \begin{scope}[shift={(0,3*\unit)}]
      \node at (0, -0.5*\unit) {\small$+$};
      \node at (0.25, -0.53*\unit) {\tiny$\cdots$};
      \node at (0.5, -0.5*\unit) {\small$+$};
      \node at (0.75, -0.5*\unit) {\small$+$};
      \node at (1, -0.53*\unit) {\tiny$\cdots$};
      \node at (1.25, -0.5*\unit) {\small$+$};
      \node at (1.5, -0.5*\unit) {\small$+$};
      \node at (2.5, -0.5*\unit) {\small$+$};
      \node at (4.75, -0.5*\unit) {\small$+$};
      \node at (5.5, -0.5*\unit) {\small$+$};
      \node at (5.75, -0.53*\unit) {\tiny$\cdots$};
      \node at (6, -0.5*\unit) {\small$+$};
    \end{scope}
    \begin{scope}[shift={(0,2*\unit)}]
      \node at (1.5, -0.5*\unit) {\small$0$};
      \node at (2.5, -0.5*\unit) {\small$0$};
      \node at (3.25, -0.5*\unit) {\small$+$};
      \node at (4.2, -0.5*\unit) {\small$+$};
      \node at (4.35, -0.53*\unit) {\tiny$\cdot$};
      \node at (4.5, -0.5*\unit) {\small$+$};
      \node at (4.75, -0.5*\unit) {\small$+$};
      \node at (5, -0.53*\unit) {\tiny$\cdots$};
      \node at (5.25, -0.5*\unit) {\small$+$};
    \end{scope}
    \begin{scope}[shift={(0,\unit)}]
      \node at (1.5, -0.5*\unit) {\small$0$};
      \node at (1.75, -0.5*\unit) {\small$+$};
      \node at (2, -0.53*\unit) {\tiny$\cdots$};
      \node at (2.25, -0.5*\unit) {\small$+$};
      \node at (2.5, -0.5*\unit) {\small$+$};
      \node at (2.75, -0.53*\unit) {\tiny$\cdots$};
      \node at (3, -0.5*\unit) {\small$+$};
      \node at (3.25, -0.5*\unit) {\small$+$};
      \node at (3.625, -0.55*\unit) {\tiny$\cdots$};
      \node at (4, -0.5*\unit) {\small$+$};
    \end{scope}
    \end{scope}
  \end{tikzpicture}
}

\newcommand*\circled[1]{%
  \tikz[baseline=(char.base)]{\node[shape=circle,draw,inner sep=1pt] (char) {$#1$};}
}

\newcommand{\lshapelabels}{%
  \begin{tikzpicture}[main node/.style={inner sep=0,minimum size =.04cm,circle,fill=black!80,draw},node distance = 0.75cm,scale=0.7]
    \draw (0,0) rectangle (0.75,2.75);
    \draw (0,0) rectangle (2.75,0.75);
    \node at (0.375, 2.375) {$+$};
    \node at (0.375, -0.375) {$\uparrow$};
    \node at (0.375, -1) {\small\text{col. }$\ell$};
    \node at (1.125, 0.375) {$+$};
    \node at (1.75, 0.34) {$\cdots$};
    \node at (2.375, 0.375) {$+$};
    \node at (2.375, -0.375) {$\uparrow$};
    \node at (2.375, -1) {\small\text{col. }$m$};
    \node at (3.25, 0.375) {$\leftarrow$};
    \node at (4.5, 0.375) {\small{row $b_B$}};
    \node at (1.25, 2.375) {$\leftarrow$};
    \node at (2.5, 2.375) {\small{row $b_T$}};
  \end{tikzpicture}
}

\newcommand{\stringplus}{%
  \begin{tikzpicture}[main node/.style={inner sep=0,minimum size =.04cm,circle,fill=black!80,draw},node distance = 0.75cm,scale=0.7]
    \draw (0,0) rectangle (2.75,0.75);
    \node at (0.375, 0.375) {$+$};
    \node at (1.375, 0.32) {$\cdots\cdots$};
    \node at (2.375, 0.375) {$+$};
  \end{tikzpicture}
}

\newcommand{\sectionrow}{%
  \begin{tikzpicture}[main node/.style={inner sep=0,minimum size =.04cm,circle,fill=black!80,draw},node distance = 0.75cm,scale=0.7]
        \draw (0.5,0) -- (0,0) -- ++(0,3.75) --++(5.75,0) -- ++(0,-0.5);
        \draw (1.5,0.5) -- ++(0,0.5) --++(3,0) --++(0,0.75) node[midway,right] {\small$b$} --++(0.5,0);
        \node[rotate=-45] at (1,0.375) {$\vdots$};
        \node[rotate=-45] at (5.25,2.5) {$\vdots$};
        \fill[pattern=north east lines,pattern color=gray!50] (1.5,1) rectangle (4.5,3.75);
        \draw[dotted] (1.5,1) rectangle (4.5,3.75);
        \node[rotate=-90] at (3,1) {$\longleftarrow$};
        \node at (3, 0.25) {\small section $b$};
    \end{tikzpicture}
}

\newcommand{\chain}{%
  \begin{tikzpicture}[main node/.style={inner sep=0,minimum size =.04cm,circle,fill=black!80,draw},node distance = 0.75cm,scale=0.7]
        \draw (0,3.75) --++(4.75,0);
        \draw (0,0) -- ++(0,0.75) --++(4.75,0) --++(0,0.45) node[midway,right] {\tiny$b$} --++(0.75,0);
        \draw[dotted] (0,0.75) rectangle (4.75,3.75);
        \fill[pattern=north east lines,pattern color=gray!50] (0,0.75) rectangle (4.75,3.75);
        \begin{scope}[scale=0.6,shift={(5.67,1.25)}]
          \lshapesmall
        \end{scope}
        \node at (3.07,1.85) {\tiny$\cdots$};
        \begin{scope}[scale=0.6,shift={(2.25,2.75)}]
          \lshapesmall
        \end{scope}
        \begin{scope}[scale=0.6,shift={(0,3.75)}]
          \stringplussmall
        \end{scope}
    \end{tikzpicture}
}

\newcommand{\lshapesmall}{%
  \draw[fill=white] (0,0) rectangle (0.625,2.25);
  \draw[fill=white] (0,0) rectangle (2.25,0.75);
  \draw (0.625,0) --++(0,0.75);
  \node at (0.32, 2) {\tiny$+$};
  \node at (1, 0.375) {\tiny$+$};
  \node at (1.5, 0.33) {\tiny$\cdots$};
  \node at (2, 0.375) {\tiny$+$};
}

\newcommand{\lvertsmall}{%
  \draw[fill=white] (0,0) rectangle (0.625,2.25);
  \draw (0,0.75) -- (0.625,0.75);
  \node at (0.32, 2) {\tiny$+$};
}

\newcommand{\stringplussmall}{%
  \draw[fill=white] (0,0) rectangle (2.25,0.75);
  \node at (0.25, 0.375) {\tiny$+$};
  \node at (1.125, 0.32) {\tiny $\cdots\cdots$};
  \node at (2, 0.375) {\tiny $+$};
}

\newcommand{\sectiononlystring}{%
  \begin{tikzpicture}[main node/.style={inner sep=0,minimum size =.04cm,circle,fill=black!80,draw},node distance = 0.75cm,scale=0.7]
        \draw (0,2.5) --++(2,0);
        \draw (0,0.25) -- ++(0,0.5) --++(2,0) --++(0,0.45) node[midway,right] {\tiny$b$} --++(0.75,0);
        \draw[dotted] (0,0.75) rectangle (2,2.5);
        \fill[pattern=north east lines,pattern color=gray!50] (0,0.75) rectangle (2,2.5);
        \draw[fill=white] (0,0.75) rectangle (2,1.2);
        \node at (0.25, 0.975) {\tiny$+$};
        \node at (1, 0.95) {\tiny $\cdots\cdots\cdots$};
        \node at (1.75, 0.975) {\tiny $+$};
    \end{tikzpicture}
}

\newcommand{\blocks}{%
  \begin{tikzpicture}[main node/.style={inner sep=0,minimum size =.04cm,circle,fill=black!80,draw},node distance = 0.75cm,scale=0.42]
    \draw (0,3) --++(2.25,0);
    \draw (0,-1) -- ++(2.25,0);
    \fill[pattern=north east lines,pattern color=gray!50] (0,-1) rectangle (2.25,3);
    \lshapesmall
    \node at (2,-1.5) {\tiny $m$};
    \node at (0.25,-1.5) {\tiny $\ell$};
    \begin{scope}[shift={(8,0.5)}]
    \draw (0,2.5) --++(2.25,0);
    \draw (0,-1.5) -- ++(2.25,0);
    \fill[pattern=north east lines,pattern color=gray!50] (0,-1.5) rectangle (2.25,2.5);
    \stringplussmall
    \node at (2,-2) {\tiny $m$};
    \node at (0.25,-2) {\tiny $\ell$};
    \end{scope}
  \end{tikzpicture}
}

\newcommand{\firstl}{%
  \begin{tikzpicture}[main node/.style={inner sep=0,minimum size =.04cm,circle,fill=black!80,draw},node distance = 0.75cm,scale=0.7]
    \draw (0,0) -- ++(0,0.5) --++(2.5,0) --++(0,0.45) node[midway,right] {\tiny$b$} --++(0.75,0);
    \begin{scope}[scale=0.6,shift={(1.92,0.833)}]
      \lshapesmall
      \node at (0.25,-0.5) {\tiny $\ell_1$};
    \end{scope}
  \end{tikzpicture}
}

\newcommand{\glueproofone}{%
  \begin{tikzpicture}[main node/.style={inner sep=0,minimum size =.04cm,circle,fill=black!80,draw},node distance = 0.75cm,scale=0.42]
    \begin{scope}[shift={(6,-1)}]
      \lshapesmall
      \node at (2,-0.5) {\tiny $m$};
      \node at (0.25,-0.5) {\tiny $\ell$};
      \node at (2.85,0.35) {\tiny $b_B$};
      \draw[thick] (1.75,0.75) --++ (0.51,0) --++(0,-0.75) --++(-0.5,0);
    \end{scope}
    \node at (9.75,0.5) {\tiny or};
    \begin{scope}[shift={(11,-1)}]
      \lvertsmall
      \node at (0.25,-0.5) {\tiny $\ell=m$};
      \node at (1.35,0.35) {\tiny $b_B$};
      \draw[thick] (0,0.75) --++ (0.635,0) --++(0,-0.75) --++(-0.635,0);
    \end{scope}
    \node at (15,0.5) {$\Longrightarrow$};
    \node at (15,1) {\small $?$};
    \begin{scope}[scale=1.2,shift={(16,-1)}]
      \lshapesmall
      \node at (1,2) {\tiny $i$};
      \node at (0.25,-0.5) {\tiny $m-1$};
      \node at (2.85,0.35) {\tiny $j$};
      \draw[thick] (-0.5,1.625) --++ (0.5,0) --++(0,-0.625) --++(-0.5,0);
      \node at (1.325,1.3125) {\tiny$\longleftarrow$};
      \node at (2,1.3125) {\tiny$b_B$};
    \end{scope}
  \end{tikzpicture}
}

\newcommand{\glueprooftwo}{%
  \begin{tikzpicture}[main node/.style={inner sep=0,minimum size =.04cm,circle,fill=black!80,draw},node distance = 0.75cm,scale=0.6]
    \draw (0.625,2.25) --++ (0,-2.25) --++(-0.625,0) --++(0,2.25);
    \draw[fill=white] (0,0) rectangle (2.25,0.625);
    \draw (0.625,0) --++(0,0.625);
    \node at (0.32, 1.75) {\tiny$+$};
    \node at (0.32, 1.25) {\tiny$0$};
    \node at (0.32, 0.75) {\small$\vert$};
    \node at (0.32, 0.315) {\tiny$0$};
    \node at (1, 0.315) {\tiny$+$};
    \node at (1.5, 0.29) {\tiny$\cdots$};
    \node at (2, 0.315) {\tiny$+$};

    \node at (0.25,-0.5) {\tiny $m-1$};
    \node at (3,0.315) {\tiny $j \geq b_B$};
    \draw[thick] (-0.5,2.06) --++ (0.5,0) --++(0,-0.625) --++(-0.5,0);
    \node at (-0.25, 1.75) {};
    \node at (1.325,1.75) {\tiny$\longleftarrow$};
    \node at (2,1.75) {\tiny$b_B$};
  \end{tikzpicture}
}

\newcommand{\doublel}{%
  \begin{tikzpicture}[main node/.style={inner sep=0,minimum size =.04cm,circle,fill=black!80,draw},node distance = 0.75cm,scale=0.6]
    \draw (0,0) rectangle (0.625,3.5);
    \draw (0,1.5) rectangle (3,2.25);
    \draw (0,0) rectangle (3,0.75);
    \draw (1.25,0) -- (1.25,0.75);
    \node[rotate=-90] at (0.92, -0.1) {\small$\longleftarrow$};
    \node at (0.93, -0.75) {\small could be $\leplus$ or $0$};
    \node at (0.32, 3.125) {\small$+$};
    \node at (1.55, 0.375) {\small$+$};
    \node at (2.15, 0.32) {\small$\cdots$};
    \node at (2.75, 0.375) {\small$+$};
    \node at (0.9, 1.875) {\small$+$};
    \node at (1.825, 1.82) {\small$\cdots\cdots$};
    \node at (2.75, 1.875) {\small$+$};
  \end{tikzpicture}
}

\newcommand{\permproofn}{%
  \begin{tikzpicture}[main node/.style={inner sep=0,minimum size =.04cm,circle,fill=black!80,draw},node distance = 0.75cm,scale=0.6]
    \draw (0,0) --++(0,0.5) node[midway,right] {\tiny $n$} --++(1.25,0);
    \draw (0,0) --++(0,1.25);
    \node at (-0.25,0.25) {\tiny $n$};
  \end{tikzpicture}
}

\newcommand{\permproofncolzero}{%
  \begin{tikzpicture}[main node/.style={inner sep=0,minimum size =.04cm,circle,fill=black!80,draw},node distance = 0.75cm,scale=0.6]
    \draw (1,2) --++(-1,0) --++(0,-2) --++(1,0);
    \draw[pattern=north east lines, pattern color=gray!50] (0,0) rectangle (0.5,2);
    \draw[fill=white] (0.75,1.25) --++(-0.75,0) --++(0,-0.5) node[midway,left] {\tiny $a_n$} --++(0.75,0);
    \draw (0.5,1.25) -- (0.5,0.75);
    \node at (0.25,-0.25) {\tiny $n$};
    \node at (0.25, 1) {\tiny $+$};
  \end{tikzpicture}
}

\newcommand{\permproofncol}{%
  \begin{tikzpicture}[main node/.style={inner sep=0,minimum size =.04cm,circle,fill=black!80,draw},node distance = 0.75cm,scale=0.6]
    \draw (2,3) --++(-2,0) --++(0,-3) --++(2,0);
    \draw[pattern=north east lines, pattern color=gray!50] (0,0) rectangle (0.5,3);
    \begin{scope}[scale=0.8,shift={(0,0.75)}]
      \lshapesmall
      \node at (0.3125, 0.375) {\tiny$+$};
      \node at (-0.5,0.375) {\tiny $a_n$};
    \end{scope}
    \node at (0.25,-0.25) {\tiny $n$};
  \end{tikzpicture}
}

\newcommand{\permprooffixedrow}{%
  \begin{tikzpicture}[main node/.style={inner sep=0,minimum size =.04cm,circle,fill=black!80,draw},node distance = 0.75cm,scale=0.6]
    \draw (0,1.75) --++(2.25,0);
    \draw (0,0) --++(1.5,0) --++(0,0.5) node[midway,right] {\tiny $i$} --++(0.75,0);
    \draw[pattern=north east lines, pattern color=gray!50] (1,0.5) rectangle (1.5,1.75);
    \draw (1,0) --++ (0,0.5);
    \node at (1.25,0.25) {\tiny $+$};
    \node at (1.25,-0.25) {\tiny $i+1$};
    \node at (1.25,2) {\tiny $i+1$};
  \end{tikzpicture}
}

\newcommand{\permprooffixedcolzero}{%
  \begin{tikzpicture}[main node/.style={inner sep=0,minimum size =.04cm,circle,fill=black!80,draw},node distance = 0.75cm,scale=0.6]
    \draw (0,1.75) --++(1.5,0);
    \draw (0,0) --++(1.5,0);
    \draw[pattern=north east lines, pattern color=gray!50] (0.25,0) rectangle (0.75,1.75);
    \draw[pattern=north east lines, pattern color=gray!50] (0.75,0) rectangle (1.25,1.75);
    \draw[fill=white] (0.25,0.625) rectangle (0.75,1.125);
    \draw[fill=white] (0.75,0.625) rectangle (1.25,1.125);
    \node at (0.5,0.875) {\tiny $+$};
    \node at (1,0.875) {\tiny $+$};
    \node at (0.5,-0.25) {\tiny $i+1$};
    \node at (1.05,-0.25) {\tiny $i$};
    \node at (0.5,2) {\tiny $i+1$};
    \node at (1.05,2) {\tiny $i$};
  \end{tikzpicture}
}

\newcommand{\permprooffixedcol}{%
  \begin{tikzpicture}[main node/.style={inner sep=0,minimum size =.04cm,circle,fill=black!80,draw},node distance = 0.75cm,scale=0.6]
    \draw (0,3) --++(1.5,0);
    \draw (0,-0.25) --++(1.5,0);
    \draw[pattern=north east lines, pattern color=gray!50] (0.25,-0.25) rectangle (0.75,3);
    \draw[pattern=north east lines, pattern color=gray!50] (0.75,-0.25) rectangle (1.25,3);
    \draw[fill=white] (0.25,1.4) rectangle (0.75,1.9) node[midway,left=4pt] {\tiny$b_L$};
    \node at (0.5,1.65) {\tiny $+$};
    \begin{scope}[scale=0.8,shift={(0.9375,0.4)}]
      \draw[fill=white] (0,0) rectangle (0.625,2.65);
      \draw[fill=white] (0,0) rectangle (2.25,0.75);
      \draw (0.625,0) --++(0,0.75);
      \node at (1, 0.375) {\tiny$+$};
      \node at (1.5, 0.33) {\tiny$\cdots$};
      \node at (2, 0.375) {\tiny$+$};
      \node at (0.3125, 0.375) {\tiny$0$};
      \node at (0.3125, 1) {\tiny$0$};
      \node at (0.3125, 1.66) {\tiny$+$};
    \end{scope}
    \draw (0.75,1.4) rectangle (1.25,1.9);
    \node at (0.5,-0.5) {\tiny $i+1$};
    \node at (1.05,-0.5) {\tiny $i$};
    \node at (0.5,3.25) {\tiny $i+1$};
    \node at (1.05,3.25) {\tiny $i$};
  \end{tikzpicture}
}

\newcommand{\permproofgeneralpath}{%
  \begin{tikzpicture}[main node/.style={inner sep=0,minimum size =.04cm,circle,fill=black!80,draw},node distance = 0.75cm,scale=0.7]
    \node at (-1,2) {\small $\hat{D}$};
    \draw[thick] (0,4.5) --++(1,0) node[midway,above] {\tiny$j=\ell_{m+1}$};
    \draw[thick] (4,0) --++(1,0) node[midway,below] {\tiny $i+1=\ell_1$};
    \fill[pattern=north east lines, pattern color=gray!50] (0.375,3.75) rectangle (0.625,4.5);
    \fill[pattern=north east lines, pattern color=gray!50] (0.75,3.625) rectangle (1.25,3.375);
    \node at (0.5,3.5) {\tiny$+$};
    \node at (1.5,3.5) {\tiny$+$};
    \fill[pattern=north east lines, pattern color=gray!50] (1.375,3.25) rectangle (1.625,3);
    \fill[pattern=north east lines, pattern color=gray!50] (1.75,2.875) rectangle (2.25,2.625);
    \node at (1.5,2.75) {\tiny$+$};
    \node at (2.5,2.25) {\tiny$\cdots$};
    \node at (3.5,1.75) {\tiny$+$};
    \fill[pattern=north east lines, pattern color=gray!50] (2.75,1.625) rectangle (3.25,1.875);
    \fill[pattern=north east lines, pattern color=gray!50] (3.375,1.5) rectangle (3.625,1.25);
    \node at (3.5,1) {\tiny$+$};
    \node at (4.5,1) {\tiny$+$};
    \fill[pattern=north east lines, pattern color=gray!50] (3.75,0.875) rectangle (4.25,1.125);
    \fill[pattern=north east lines, pattern color=gray!50] (4.375,0.75) rectangle (4.625,0);
    \node at (3.5,-0.30) {\tiny$\ell_2$};
    \node at (2.5,-0.30) {\tiny$\cdots$};
    \node at (1.5,-0.30) {\tiny$\ell_m$};
    \node at (6.0,1) {\tiny$b_1$};
    \node at (6.0,1.75) {\tiny$b_2$};
    \node[rotate=90] at (6.05,2.25) {\tiny$\cdots$};
    \node at (6.0,2.75) {\tiny$b_{m-1}$};
    \node at (6.0,3.5) {\tiny$b_{m}$};
    \draw[->,decorate,decoration=snake] (8,2) -- node[above] {\tiny path $\hat\pi(i+1)$} (10,2);
    \begin{scope}[shift={(11,0)}]
      \node[main node, label=below:{\tiny$i+1$}] (i) at (3.5,0) {};
      \node[main node, label=above:{\tiny$j$}] (j) at (0,4.5) {};
      \draw[rounded corners,postaction={decorate,decoration={markings,mark=at position 0.1 with {\arrow[scale=0.75]{>}}}}] (i) |- (2.5,1) |- (2,1.75);
      \node[rotate=-45] at (1.75,2.2) {\tiny$\cdots$};
      \draw[rounded corners,postaction={decorate,decoration={markings,mark=at position 0.9 with {\arrow[scale=0.75]{>}}}}] (1.5,2.75) -| (1,3.5) -| (j);
    \end{scope}
  \end{tikzpicture}
}

\newcommand{\permproofihat}{%
  \begin{tikzpicture}[main node/.style={inner sep=0,minimum size =.04cm,circle,fill=black!80,draw},node distance = 0.75cm,scale=0.7]
    \draw (-1,0.25) rectangle (-0.5,4.5);
    \node at (-0.75,-0.30) {\tiny $a_n$};
    \draw[thick] (0,4.5) --++(1,0) node[midway,above] {\tiny$j=\ell_{m+1}$};
    \node at (4.5,-0.30) {\tiny $\ell_1$};
    \node at (0.5,3.5) {\tiny$+$};
    \node[draw,circle,inner sep=1pt] at (1.5,3.5) {\tiny$+$};
    \node at (1.5,2.75) {\tiny$+$};
    \node at (2.5,2.25) {\tiny$\cdots$};
    \node[draw,circle,inner sep=1pt] at (3.5,1.75) {\tiny$+$};
    \node at (3.5,1) {\tiny$+$};
    \node[draw,circle,inner sep=1pt] at (4.5,1) {\tiny$+$};
    \node at (3.5,-0.30) {\tiny$\ell_2$};
    \node at (2.5,-0.30) {\tiny$\cdots$};
    \node at (1.5,-0.30) {\tiny$\ell_m$};
    \node at (6.0,1) {\tiny$b_1$};
    \node at (6.0,1.75) {\tiny$b_2$};
    \node[rotate=90] at (6.05,2.25) {\tiny$\cdots$};
    \node at (6.0,2.75) {\tiny$b_{m-1}$};
    \node at (6.0,3.5) {\tiny$b_{m}$};
    \node at (7.5,2.5) {\small (i)};
    \node at (8.5,2.5) {\small or};
    \node at (8.5,1.5) {\small $\hat{D}$};
    \begin{scope}[shift={(11,0)}]
    \node at (-1.5,2.5) {\small (iv)};
    \draw[dashed] (0,5.25) --++ (5,0)  node[right=4pt] {\tiny $a_n$};
    \draw[thick] (-0.5,3.75) --++(0,1) node[midway,left] {\tiny$j$};
    \node at (0.5,4.25) {\tiny$+$};
    \node at (0.5,3.5) {\tiny$+$};
    \node[draw,circle,inner sep=1pt] at (1.5,3.5) {\tiny$+$};
    \node at (1.5,2.75) {\tiny$+$};
    \node at (2.5,2.25) {\tiny$\cdots$};
    \node[draw,circle,inner sep=1pt] at (3.5,1.75) {\tiny$+$};
    \node at (3.5,1) {\tiny$+$};
    \node[draw,circle,inner sep=1pt] at (4.5,1) {\tiny$+$};
    \node at (4.5,-0.30) {\tiny$\ell_1$};
    \node at (3.5,-0.30) {\tiny$\ell_2$};
    \node at (2.5,-0.30) {\tiny$\cdots$};
    \node at (1.5,-0.30) {\tiny$\ell_m$};
    \node at (0.5,-0.30) {\tiny$\ell_{m+1}$};
    \node at (6.0,1) {\tiny$b_1$};
    \node at (6.0,1.75) {\tiny$b_2$};
    \node[rotate=90] at (6.05,2.25) {\tiny$\cdots$};
    \node at (6.0,2.75) {\tiny$b_{m-1}$};
    \node at (6.0,3.5) {\tiny$b_{m}$};
    \end{scope}
  \end{tikzpicture}
}

\newcommand{\permprooficols}{%
  \begin{tikzpicture}[main node/.style={inner sep=0,minimum size =.04cm,circle,fill=black!80,draw},node distance = 0.75cm,scale=0.7]
    \node at (2,1) {\small $\hat{D}$};
    \node[draw,circle,inner sep=1pt] at (3.5,2) {\tiny$+$};
    \node at (3.5,1) {\tiny$+$};
    \node[draw,circle,inner sep=1pt] at (4.5,1) {\tiny$+$};
    \node at (4.5,0.25) {\tiny$+$};
    \node at (4.5,-0.75) {\tiny$\ell_s$};
    \node at (3.5,-0.75) {\tiny$\ell_{s+1}$};
    \node at (6.0,0.25) {\tiny$b_{s-1}$};
    \node at (6.0,1) {\tiny$b_s$};
    \node at (6.0,2) {\tiny$b_{s+1}$};
    \node at (8.25,1) {$\longmapsto$};
    \begin{scope}[shift={(8,0)}]
      \node at (2,1) {\small ${D}$};
      \draw (3.25,1) -- (3.25,2.5);
      \draw (3.75,1) -- (3.75,2.5);
      \draw (4.25,0.5) -- (4.25,2.5);
      \draw (4.75,0.5) -- (4.75,2.5);
      \fill[pattern=north east lines, pattern color=gray!50] (3.75,1.75) rectangle (4.25,2.25);
      \fill[pattern=north east lines, pattern color=gray!50] (4.25,1.75) rectangle (4.75,1.25);
      \node[draw,circle,inner sep=1pt] at (3.5,2) {\tiny$+$};
      \node at (4.5,2) {\tiny$+$};
      \node[draw,circle,inner sep=1pt] at (4.5,1) {\tiny$+$};
      \node at (4.5,-0.5) {\tiny$\ell_s$};
      \node at (3.5,-0.5) {\tiny$\ell_{s+1}$};
      \node at (6.0,1) {\tiny$b_s$};
      \node at (6.0,2) {\tiny$b_{s+1}$};
    \end{scope}
  \end{tikzpicture}
}

\newcommand{\permproofi}{%
  \begin{tikzpicture}[main node/.style={inner sep=0,minimum size =.04cm,circle,fill=black!80,draw},node distance = 0.75cm,scale=0.7]
    \draw[thick] (0,4.5) --++(1,0) node[midway,above] {\tiny$j=\ell_{m+1}$};
    \draw (0.25,3) --++(0,0.75) --++(0.5,0) --++ (0,-0.75);
    \fill[pattern=north east lines, pattern color=gray!50] (0.25,3.75) rectangle (0.75,4.5);
    \fill[pattern=north east lines, pattern color=gray!50] (0.75,3.75) rectangle (1.25,3.25);
    \node at (0.5,3.5) {\tiny$+$};
    \draw (1.25,4) -- (1.25,2.25);
    \draw (1.75,4) -- (1.75,2.25);
    \fill[pattern=north east lines, pattern color=gray!50] (1.25,3.25) rectangle (1.75,3);
    \fill[pattern=north east lines, pattern color=gray!50] (1.75,3) rectangle (2.25,2.5);
    \node at (1.5,3.5) {\tiny$+$};
    \node[draw,circle,inner sep=1pt] at (1.5,2.75) {\tiny$+$};
    \node at (2.5,2.25) {\tiny$\cdots$};
    \draw (3.25,2.25) -- (3.25,0.5);
    \draw (3.75,2.25) -- (3.75,0.5);
    \fill[pattern=north east lines, pattern color=gray!50] (2.75,1.5) rectangle (3.25,2);
    \fill[pattern=north east lines, pattern color=gray!50] (3.25,1.5) rectangle (3.75,1.25);
    \node at (3.5,1.75) {\tiny$+$};
    \node[draw,circle,inner sep=1pt] at (3.5,1) {\tiny$+$};
    \draw (4.25,1.5) -- (4.25,-0.25);
    \draw (4.75,1.5) -- (4.75,-0.25);
    \fill[pattern=north east lines, pattern color=gray!50] (3.75,0.75) rectangle (4.25,1.25);
    \fill[pattern=north east lines, pattern color=gray!50] (4.25,0.75) rectangle (4.75,0.5);
    \node at (4.5,1) {\tiny$+$};
    \node[draw,circle,inner sep=1pt] at (4.5,0.25) {\tiny$+$};
    \node at (4.5,-1) {\tiny$\ell_1$};
    \node at (3.5,-1) {\tiny$\ell_2$};
    \node at (2.5,-1) {\tiny$\cdots$};
    \node at (1.5,-1) {\tiny$\ell_m$};
    \node at (6.0,0.25) {\tiny$b_1$};
    \node at (6.0,1) {\tiny$b_2$};
    \node at (6.0,1.75) {\tiny$b_3$};
    \node[rotate=90] at (6.05,2.25) {\tiny$\cdots$};
    \node at (6.0,2.75) {\tiny$b_{m}$};
    \node at (6.0,3.5) {\tiny$b$};
    \node at (7.5,2.5) {\small (i)};
    \node at (8.5,2.5) {\small or};
    \node at (8.5,1.5) {\small ${D}$};
    \begin{scope}[shift={(11,0)}]
      \node at (-1.5,2.5) {\small (iv)};
      \draw[dashed] (-0.5,4.5) --++ (6,0)  node[right=4pt] {\tiny $a_n$};
      \draw[thick] (-0.5,3) --++(0,1);
      \node at (-0.75,3.5) {\tiny$j$};
      \draw (0.25,3) -- (0.25,4);
      \draw (0.75,3) -- (0.75,4);
      \fill[pattern=north east lines, pattern color=gray!50] (0.25,3.75) rectangle (-0.5,3.25);
      \fill[pattern=north east lines, pattern color=gray!50] (0.75,3.75) rectangle (1.25,3.25);
      \node at (0.5,3.5) {\tiny$0$};
      \draw (1.25,4) -- (1.25,2.25);
      \draw (1.75,4) -- (1.75,2.25);
      \fill[pattern=north east lines, pattern color=gray!50] (1.25,3.25) rectangle (1.75,3);
      \fill[pattern=north east lines, pattern color=gray!50] (1.75,3) rectangle (2.25,2.5);
      \node at (1.5,3.5) {\tiny$+$};
      \node[draw,circle,inner sep=1pt] at (1.5,2.75) {\tiny$+$};
      \node at (2.5,2.25) {\tiny$\cdots$};
      \draw (3.25,2.25) -- (3.25,0.5);
      \draw (3.75,2.25) -- (3.75,0.5);
      \fill[pattern=north east lines, pattern color=gray!50] (2.75,1.5) rectangle (3.25,2);
      \fill[pattern=north east lines, pattern color=gray!50] (3.25,1.5) rectangle (3.75,1.25);
      \node at (3.5,1.75) {\tiny$+$};
      \node[draw,circle,inner sep=1pt] at (3.5,1) {\tiny$+$};
      \draw (4.25,1.5) -- (4.25,-0.25);
      \draw (4.75,1.5) -- (4.75,-0.25);
      \fill[pattern=north east lines, pattern color=gray!50] (3.75,0.75) rectangle (4.25,1.25);
      \fill[pattern=north east lines, pattern color=gray!50] (4.25,0.75) rectangle (4.75,0.5);
      \node at (4.5,1) {\tiny$+$};
      \node[draw,circle,inner sep=1pt] at (4.5,0.25) {\tiny$+$};
      \node at (4.5,-1) {\tiny$\ell_1$};
      \node at (3.5,-1) {\tiny$\ell_2$};
      \node at (2.5,-1) {\tiny$\cdots$};
      \node at (1.5,-1) {\tiny$\ell_m$};
      \node at (0.5,-1) {\tiny$\ell_{m+1}$};
      \node at (6.0,0.25) {\tiny$b_1$};
      \node at (6.0,1) {\tiny$b_2$};
      \node at (6.0,1.75) {\tiny$b_3$};
      \node[rotate=90] at (6.05,2.25) {\tiny$\cdots$};
      \node at (6.0,2.75) {\tiny$b_{m}$};
      \node at (6.0,3.5) {\tiny$b=j$};
    \end{scope}
  \end{tikzpicture}
}

\newcommand{\permproofibothcols}{%
  \begin{tikzpicture}[main node/.style={inner sep=0,minimum size =.04cm,circle,fill=black!80,draw},node distance = 0.75cm,scale=0.7]
    \draw (0,0) --++(1.5,0);
    \draw (0.25,1.5) --++(0,-0.875) --++(0.5,0) --++(0,0.875);
    \fill[pattern=north east lines, pattern color=gray!50] (0.75,0) rectangle (1.25,0.625);
    \node[circle,inner sep=1pt,draw] at (0.5,0.875) {\tiny $+$};
    \node at (1,0.875) {\tiny $+$};
    \node at (0,-0.5) {\tiny $\ell_1 = i+1$};
    \node at (1,-0.5) {\tiny $i$};
    \node at (-0.25,0.875) {\tiny $b_1$};
    \draw[->,decorate,decoration={snake,amplitude=2pt}] (2.25,0.5) -- (2.75,0.5);
    \begin{scope}[shift={(3.25,0)}]
      \node[main node, label=below:{\tiny$i$}] (i) at (1,0) {};
      \node[circle,inner sep=1pt,draw] at (0.5,0.875) {};
      \node[main node, label=above:{\tiny$\ell_1$}] (j) at (0.5,0.875) {};
      \draw[rounded corners,postaction={decorate,decoration={markings,mark=at position 0.25 with {\arrow[scale=0.75]{>}}}}] (i) |- (j);
    \end{scope}
  \end{tikzpicture}
}

\newcommand{\permprooficolrow}{%
  \begin{tikzpicture}[main node/.style={inner sep=0,minimum size =.04cm,circle,fill=black!80,draw},node distance = 0.75cm,scale=0.7]
    \draw (0,0) --++(0.75,0) --++(0,0.75) node[midway,right] {\tiny $i = b_1$};
    \node[circle,inner sep=1pt,draw] at (0.375,0.375) {\tiny $+$};
    \node at (0,-0.25) {\tiny $\ell_1 = i+1$};
    \draw[->,decorate,decoration={snake,amplitude=2pt}] (2.25,0.375) -- (2.75,0.375);
    \begin{scope}[shift={(3.25,0)}]
      \node[main node, label=right:{\tiny$i$}] (i) at (1.25,0.375) {};
      \node[circle,inner sep=1pt,draw] at (0.5,0.375) {};
      \node[main node, label=above:{\tiny$\ell_1$}] (j) at (0.5,0.375) {};
      \draw[postaction={decorate,decoration={markings,mark=at position 0.5 with {\arrow[scale=0.75]{>}}}}] (i) -- (j);
    \end{scope}
  \end{tikzpicture}
}

\newcommand{\permproofibothrows}{%
  \begin{tikzpicture}[main node/.style={inner sep=0,minimum size =.04cm,circle,fill=black!80,draw},node distance = 0.75cm,scale=0.7]
    \draw (1.5,0) --++(0,1.5);
    \draw (0.125,1.5) -- (0.125,0.25);
    \draw (0.625,1.5) -- (0.625,0.25);
    \fill[pattern=north east lines, pattern color=gray!50] (0.625,0.625) rectangle (1.5,1.125);
    \node[circle,inner sep=1pt,draw] at (0.375,0.875) {\tiny $+$};
    \node at (2,0.25) {\tiny $i+1$};
    \node at (2,0.875) {\tiny $i = b_1$};
    \node at (0.375,-0.25) {\tiny $\ell_1$};
    \draw[->,decorate,decoration={snake,amplitude=2pt}] (3,0.875) -- (3.5,0.875);
    \begin{scope}[shift={(4,0.5)}]
      \node[main node, label=right:{\tiny$i$}] (i) at (1.25,0.375) {};
      \node[circle,inner sep=1pt,draw] at (0.5,0.375) {};
      \node[main node, label=above:{\tiny$\ell_1$}] (j) at (0.5,0.375) {};
      \draw[postaction={decorate,decoration={markings,mark=at position 0.5 with {\arrow[scale=0.75]{>}}}}] (i) -- (j);
    \end{scope}
  \end{tikzpicture}
}

\newcommand{\permproofirowcol}{%
  \begin{tikzpicture}[main node/.style={inner sep=0,minimum size =.04cm,circle,fill=black!80,draw},node distance = 0.75cm,scale=0.7]
    \draw (1.25,-1.25) --++(0,1) --++(1,0);
    \draw (0.125,1.5) -- (0.125,0.25);
    \draw (0.625,1.5) -- (0.625,0.25);
    \fill[pattern=north east lines, pattern color=gray!50] (0.625,0.625) rectangle (1.5,1.125);
    \fill[pattern=north east lines, pattern color=gray!50] (1.5,0.625) rectangle (2,-0.25);
    \node[circle,inner sep=1pt,draw] at (0.375,0.875) {\tiny $+$};
    \node at (1.75,0.875) {\tiny $+$};
    \node at (1.75,-0.5) {\tiny $i$};
    \node at (1.75,-1) {\tiny $i+1$};
    \node at (0.375,2) {\tiny $\ell_1$};
    \node at (-0.5,0.875) {\tiny $b_1$};
    \draw[->,decorate,decoration={snake,amplitude=2pt}] (3,0.25) -- (3.5,0.25);
    \begin{scope}[shift={(4,0)}]
      \node[main node, label=below:{\tiny$i$}] (i) at (1.75,-0.25) {};
      \node[circle,inner sep=1pt,draw] at (0.375,0.875) {};
      \node[main node, label=above:{\tiny$\ell_1$}] (j) at (0.375,0.875) {};
      \draw[rounded corners,postaction={decorate,decoration={markings,mark=at position 0.25 with {\arrow[scale=0.75]{>}}}}] (i) |- (j);
    \end{scope}
  \end{tikzpicture}
}

\newcommand{\permproofiihat}{%
  \begin{tikzpicture}[main node/.style={inner sep=0,minimum size =.04cm,circle,fill=black!80,draw},node distance = 0.75cm,scale=0.7]
    \draw[thick] (0,6.5) --++(1,0) node[midway,above] {\tiny$\ell_{m+1}$};
    \node at (0.5,5.25) {\tiny$+$};
    \node at (1.5,5.25) {\tiny$+$};
    \node at (1.5,4.5) {\tiny$+$};
    \node at (2.25,4) {\tiny$\cdots$};
    \node at (3,3.5) {\tiny$+$};
    \node at (3,2.75) {\tiny$+$};
    \node at (3.75,2.25) {\tiny$\cdots$};
    \node[draw,circle,inner sep=1pt] at (4.5,1.75) {\tiny$+$};
    \node at (4.5,1) {\tiny$+$};
    \node[draw,circle,inner sep=1pt] at (5.5,1) {\tiny$+$};
    \node at (5.5,-0.30) {\tiny$\ell_1$};
    \node at (4.5,-0.30) {\tiny$\ell_2$};
    \node at (3.85,-0.30) {\tiny$\cdots$};
    \node at (3,-0.30) {\tiny$a_n = \ell_s$};
    \draw (2.75,0.5) rectangle (3.25,6.5);
    \node at (3,6.25) {\tiny$*$};
    \node at (3,5.25) {\tiny$*$};
    \node at (3,4.5) {\tiny$*$};
    \node at (2.15,-0.30) {\tiny$\cdots$};
    \node at (1.5,-0.30) {\tiny$\ell_m$};
    \node at (7.0,1) {\tiny$b_1$};
    \node at (7.0,1.75) {\tiny$b_2$};
    \node[rotate=90] at (7.05,2.25) {\tiny$\cdots$};
    \node at (7.0,2.75) {\tiny$b_{s-1}$};
    \node at (7.0,3.5) {\tiny$b_{s}$};
    \node[rotate=90] at (7.05,4) {\tiny$\cdots$};
    \node at (7.0,4.5) {\tiny$b_{m-1}$};
    \node at (7.0,5.25) {\tiny$b_{m}$};
    \node[rotate=90] at (7.05,5.75) {\tiny$\cdots$};
    \node at (7.0,6.25) {\tiny$1$};
    \draw[decorate,decoration={calligraphic brace,amplitude=6pt,raise=5pt}] (7.25,6) -- (7.25,2.75) node [midway, right=20pt,align=center] {\tiny rows with a $\leplus$ \\ \tiny in column $a_n$};
    \node at (8.5,2.5) {\small (ii)};
    \node at (9.5,2.5) {\small or};
    \node at (9.5,1.5) {\small $\hat{D}$};
    \begin{scope}[shift={(12,0)}]
    \node at (-1.5,2.5) {\small (iii)};
    \draw[thick] (0,6.5) --++(1,0) node[midway,above] {\tiny$\ell_{m+1}$};
    \node at (0.5,5.25) {\tiny$+$};
    \node at (1.5,5.25) {\tiny$+$};
    \node at (1.5,4.5) {\tiny$+$};
    \node at (2.25,4.062) {\tiny$\cdots$};
    \node at (3,3.625) {\tiny$+$};
    \draw[dashed] (0,3.125) --++ (6.5,0)  node[right=3pt] {\tiny $a_n$};
    \node at (3,2.625) {\tiny$+$};
    \node at (3.75,2.187) {\tiny$\cdots$};
    \node[draw,circle,inner sep=1pt] at (4.5,1.75) {\tiny$+$};
    \node at (4.5,1) {\tiny$+$};
    \node[draw,circle,inner sep=1pt] at (5.5,1) {\tiny$+$};
    \node at (5.5,-0.30) {\tiny$\ell_1$};
    \node at (4.5,-0.30) {\tiny$\ell_2$};
    \node at (3.85,-0.30) {\tiny$\cdots$};
    \node at (3,-0.30) {\tiny$\ell_s$};
    \node at (2.25,-0.30) {\tiny$\cdots$};
    \node at (1.5,-0.30) {\tiny$\ell_m$};
    \node at (7.0,1) {\tiny$b_1$};
    \node at (7.0,1.75) {\tiny$b_2$};
    \node[rotate=90] at (7.05,2.187) {\tiny$\cdots$};
    \node at (7.0,2.625) {\tiny$b_{s-1}$};
    \node at (7.0,3.625) {\tiny$b_{s}$};
    \node[rotate=90] at (7.05,4.062) {\tiny$\cdots$};
    \node at (7.0,4.5) {\tiny$b_{m-1}$};
    \node at (7.0,5.25) {\tiny$b_{m}$};
    \draw[decorate,decoration={calligraphic brace,mirror,amplitude=6pt,raise=4pt}] (0,6) -- (0,3.5);
    \end{scope}
  \end{tikzpicture}
}

\newcommand{\permproofii}{%
  \begin{tikzpicture}[main node/.style={inner sep=0,minimum size =.04cm,circle,fill=black!80,draw},node distance = 0.75cm,scale=0.7]
    \draw[thick] (0,4.5) --++(1,0) node[midway,above] {\tiny$\ell_{m+1}$};
    \draw (0.25,3) --++(0,1);
    \draw (0.75,4) --++ (0,-1);
    \fill[pattern=north east lines, pattern color=gray!50] (0.75,3.75) rectangle (1.25,3.25);
    \node at (0.5,3.5) {\tiny$+$};
    \node at (1.25,3) {\tiny$\cdots$};
    \draw (1.75,3) -- (1.75,1.25);
    \draw (2.25,3) -- (2.25,1.25);
    \fill[pattern=north east lines, pattern color=gray!50] (1.75,2.25) rectangle (1.25,2.75);
    \fill[pattern=north east lines, pattern color=gray!50] (1.75,2.25) rectangle (2.25,2);
    \fill[pattern=north east lines, pattern color=gray!50] (2.25,2) rectangle (2.75,1.5);
    \node at (2,2.5) {\tiny$+$};
    \node at (2,1.75) {\tiny$+$};
    \draw (2.75,2.25) -- (2.75,0.5);
    \draw (3.25,2.25) -- (3.25,0.5);
    \fill[pattern=north east lines, pattern color=gray!50] (2.75,1.5) rectangle (3.25,1.25);
    \fill[pattern=north east lines, pattern color=gray!50] (3.25,0.75) rectangle (3.75,1.25);
    \node at (3,1.75) {\tiny$+$};
    \node[draw,circle,inner sep=1pt] at (3,1) {\tiny$+$};
    \node at (3.75,0.5) {\tiny$\cdots$};
    \draw (4.25,0.5) -- (4.25,-1.25);
    \draw (4.75,0.5) -- (4.75,-1.25);
    \fill[pattern=north east lines, pattern color=gray!50] (3.75,-0.25) rectangle (4.25,0.25);
    \fill[pattern=north east lines, pattern color=gray!50] (4.25,-0.25) rectangle (4.75,-0.5);
    \node at (4.5,0) {\tiny$+$};
    \node[draw,circle,inner sep=1pt] at (4.5,-0.75) {\tiny$+$};
    \node at (4.5,-1.750) {\tiny$\ell_1$};
    \node at (3.75,-1.750) {\tiny$\cdots$};
    \node at (3,-1.750) {\tiny$\ell_{s-1}$};
    \node at (2,-1.750) {\tiny$\ell_{s+1}$};
    \node at (1.25,-1.750) {\tiny$\cdots$};
    \node at (0.5,-1.750) {\tiny$\ell_{m+1}$};
    \draw[decorate,decoration={calligraphic brace,mirror,amplitude=6pt,raise=3pt}] (0.25,-2) -- (2.25,-2) node[midway,below=12pt] {\tiny original path};
    \draw[decorate,decoration={calligraphic brace,mirror,amplitude=6pt,raise=3pt}] (2.75,-2) -- (4.75,-2) node[midway,below=12pt] {\tiny same as (i), (iv)};
    \node at (6.0,-0.75) {\tiny$b_1$};
    \node at (6.0,0) {\tiny$b_2$};
    \node[rotate=90] at (6.05,0.5) {\tiny$\cdots$};
    \node at (6.0,1) {\tiny$b_{s-1}$};
    \node at (6.0,1.75) {\tiny$b_{s}$};
    \node at (6.0,2.5) {\tiny$b_{s+1}$};
    \node[rotate=90] at (6.05,3) {\tiny$\cdots$};
    \node at (6.0,3.5) {\tiny$b_m$};
    \node at (7.5,2) {\small (ii)};
    \node at (8.5,2) {\small or};
    \node at (8.5,1) {\small $D$};
    \begin{scope}[shift={(11.5,-0.75)}]
    \node at (-2,2.75) {\small (iii)};
    \draw[thick] (-1,5.25) --++(1,0) node[midway,above] {\tiny$\ell_{m+1}$};
    \draw (-0.75,3.75) --++(0,1);
    \draw (-0.25,4.75) --++ (0,-1);
    \fill[pattern=north east lines, pattern color=gray!50] (-0.25,4.5) rectangle (0.25,4);
    \node at (-0.5,4.25) {\tiny$+$};
    \node at (0.25,3.75) {\tiny$\cdots$};
    \draw (0.75,3.75) -- (0.75,2);
    \draw (1.25,3.75) -- (1.25,2);
    \fill[pattern=north east lines, pattern color=gray!50] (0.75,3) rectangle (0.25,3.5);
    \fill[pattern=north east lines, pattern color=gray!50] (0.75,3) rectangle (1.25,2.75);
    \fill[pattern=north east lines, pattern color=gray!50] (1.25,2.75) rectangle (1.75,2.25);
    \node at (1,3.25) {\tiny$+$};
    \node at (1,2.5) {\tiny$+$};
    \draw (1.75,3) -- (1.75,1.25);
    \draw (2.25,3) -- (2.25,1.25);
    \fill[pattern=north east lines, pattern color=gray!50] (1.75,2.25) rectangle (2.25,2);
    \fill[pattern=north east lines, pattern color=gray!50] (2.25,2) rectangle (2.75,1.5);
    \node at (2,2.5) {\tiny$+$};
    \node at (2,1.75) {\tiny$+$};
    \draw (2.75,2.25) -- (2.75,0.5);
    \draw (3.25,2.25) -- (3.25,0.5);
    \fill[pattern=north east lines, pattern color=gray!50] (2.75,1.5) rectangle (3.25,1.25);
    \fill[pattern=north east lines, pattern color=gray!50] (3.25,0.75) rectangle (3.75,1.25);
    \node at (3,1.75) {\tiny$+$};
    \node[draw,circle,inner sep=1pt] at (3,1) {\tiny$+$};
    \node at (3.75,0.5) {\tiny$\cdots$};
    \draw (4.25,0.5) -- (4.25,-1.25);
    \draw (4.75,0.5) -- (4.75,-1.25);
    \fill[pattern=north east lines, pattern color=gray!50] (3.75,-0.25) rectangle (4.25,0.25);
    \fill[pattern=north east lines, pattern color=gray!50] (4.25,-0.25) rectangle (4.75,-0.5);
    \node at (4.5,0) {\tiny$+$};
    \node[draw,circle,inner sep=1pt] at (4.5,-0.75) {\tiny$+$};
    \node at (4.5,-1.750) {\tiny$\ell_1$};
    \node at (3.75,-1.750) {\tiny$\cdots$};
    \node at (3,-1.750) {\tiny$\ell_{s-1}$};
    \node at (2,-1.750) {\tiny$\ell_{s}$};
    \node at (1,-1.750) {\tiny$\ell_{s+1}$};
    \node at (0.25,-1.750) {\tiny$\cdots$};
    \node at (-0.5,-1.750) {\tiny$\ell_{m+1}$};
    \draw[decorate,decoration={calligraphic brace,amplitude=6pt,raise=3pt}] (6.25,5) -- (6.25,2.5) node[midway,right=12pt] {\tiny original path};
    \draw[decorate,decoration={calligraphic brace,amplitude=6pt,raise=3pt}] (6.25,1) -- (6.25,-0.75) node[midway,right=12pt] {\tiny same as (i), (iv)};
    \node at (6.0,-0.75) {\tiny$b_1$};
    \node at (6.0,0) {\tiny$b_2$};
    \node[rotate=90] at (6.05,0.5) {\tiny$\cdots$};
    \node at (6.0,1) {\tiny$b_{s-1}$};
    \node at (6.0,1.75) {\tiny$a_{n}$};
    \node at (6.0,2.5) {\tiny$b_{s}$};
    \node at (6.0,3.25) {\tiny$b_{s+1}$};
    \node[rotate=90] at (6.05,3.75) {\tiny$\cdots$};
    \node at (6.0,4.25) {\tiny$b_m$};
    \end{scope}
  \end{tikzpicture}
}

\newcommand{\permproofiiispecialone}{%
  \begin{tikzpicture}[main node/.style={inner sep=0,minimum size =.04cm,circle,fill=black!80,draw},node distance = 0.75cm,scale=0.7]
    \node at (2,1) {\small $\hat{D}$};
    \draw[thick] (3,2.5) --++(1,0) node[midway,above] {\tiny$j=\ell_{m+1}$};
    \draw[dashed] (3,1.75) --++ (2.5,0) node[right=3pt] {\tiny $a_n$};
    \node at (3.5,1) {\tiny$+$};
    \node[draw,circle,inner sep=1pt] at (4.5,1) {\tiny$+$};
    \node at (4.5,0.25) {\tiny$+$};
    \node at (4.5,-0.75) {\tiny$\ell_m$};
    \node at (6.0,0.25) {\tiny$b_{m-1}$};
    \node at (6.0,1) {\tiny$b_m$};
    \node at (8.25,1) {$\longmapsto$};
    \begin{scope}[shift={(8,0)}]
      \node at (2,1) {\small ${D}$};
      \draw[thick] (3,2.5) --++(1,0) node[midway,above] {\tiny$j=\ell_{m+1}$};
      \draw (3.25,1) -- (3.25,2) --++ (0.5,0) -- (3.75,1);
      \draw (4.25,0.5) -- (4.25,2.15);
      \draw (4.75,0.5) -- (4.75,2.15);
      \fill[pattern=north east lines, pattern color=gray!50] (3.25,2) rectangle (3.75,2.5);
      \fill[pattern=north east lines, pattern color=gray!50] (3.75,1.5) rectangle (4.25,2);
      \fill[pattern=north east lines, pattern color=gray!50] (4.25,1.5) rectangle (4.75,1.25);
      \node at (3.5,1.75) {\tiny$+$};
      \node at (4.5,1.75) {\tiny$+$};
      \node[draw,circle,inner sep=1pt] at (4.5,1) {\tiny$+$};
      \node at (4.5,-0.5) {\tiny$\ell_m$};
      \node at (6.0,1) {\tiny$b_m$};
      \node at (6.0,1.75) {\tiny$a_{n}$};
    \end{scope}
  \end{tikzpicture}
}

\newcommand{\permproofiiispecialtwo}{%
  \begin{tikzpicture}[main node/.style={inner sep=0,minimum size =.04cm,circle,fill=black!80,draw},node distance = 0.75cm,scale=0.7]
    \node at (2,0) {\small $\hat{D}$};
    \draw[thick] (4,-1.25) --++(1,0) node[midway,below] {\tiny$i+1=\ell_{1}$};
    \draw[dashed] (3,-0.5) --++ (2.5,0) node[right=3pt] {\tiny $a_n$};
    \node at (3.5,1) {\tiny$+$};
    \node at (3.5,0.25) {\tiny$+$};
    \node at (4.5,0.25) {\tiny$+$};
    \node at (3.5,-1.58) {\tiny$\ell_2$};
    \node at (6.0,0.25) {\tiny$b_{1}$};
    \node at (6.0,1) {\tiny$b_2$};
    \node at (8.25,0) {$\longmapsto$};
    \begin{scope}[shift={(8,0)}]
      \node at (2,0) {\small ${D}$};
      \draw[thick] (4,-1.25) --++(1,0) node[midway,below] {\tiny$i+1=\ell_{1}$};
      \node at (6,-0.5){\tiny $a_n$};
      \node at (3.5,1) {\tiny$+$};
      \node at (3.5,0.25) {\tiny$+$};
      \node at (4.5,0.25) {\tiny$+$};
      \node at (4.5,-0.5) {\tiny$+$};
      \draw (3.25,1.5) -- (3.25,-0.25);
      \draw (3.75,1.5) -- (3.75,-0.25);
      \draw (4.25,0.75) -- (4.25,-0.75) --++(0.5,0) -- (4.75,0.75);
      \fill[pattern=north east lines, pattern color=gray!50] (3.25,0.5) rectangle (3.75,0.75);
      \fill[pattern=north east lines, pattern color=gray!50] (3.75,0) rectangle (4.25,0.5);
      \fill[pattern=north east lines, pattern color=gray!50] (4.25,0) rectangle (4.75,-0.25);
      \node at (3.5,-1.58) {\tiny$\ell_2$};
      \node at (6.0,0.25) {\tiny$b_{1}$};
      \node at (6.0,1) {\tiny$b_2$};
    \end{scope}
  \end{tikzpicture}
}

\newcommand{\permproofpathex}{%
  \begin{tikzpicture}[main node/.style={inner sep=0,minimum size =.04cm,circle,fill=black!80,draw},node distance = 0.75cm,scale=0.7]
    \pgfmathsetmacro{\unit}{0.75};
    \draw (0,5*\unit) --++ (7*\unit,0) --++(0,-1*\unit) node[main node,midway,label=right:{\tiny (i)}] (i) {} --++(-1*\unit,0) --++(0,-1*\unit) --++(-1*\unit,0) node[main node,midway,label=below:{\tiny (i)}] (i2) {} --++(0,-1*\unit) --++(-1*\unit,0) node[main node,midway, label=below:{\tiny (ii)}] (ii) {} --++(0,-1*\unit) node[main node,midway] (ii2) {} --++(-3*\unit,0) --++(0,-1*\unit);
    \node[main node] (ib) at (4.875,3.75) {};
    \node[main node] (i2b) at (2.625,3.75) {};
    \node[main node] (iib) at (-0.25,2.75) {};
    \node[main node] (ii2b) at (1.125,3.75) {};
    \draw (1.625,0.5) rectangle (2.125,4);
    \node at (1.875, 0.15) {\tiny $a_n$};
    \draw[rounded corners,postaction={decorate,decoration={markings,mark=at position 0.8 with {\arrow[scale=0.75]{>}}}}] (i) -| (ib);
    \draw[rounded corners,postaction={decorate,decoration={markings,mark=at position 0.25 with {\arrow[scale=0.75]{>}},mark=at position 0.95 with {\arrow[scale=0.75]{>}}}}] (i2) |- (3.325,2.625) |- (2.625,3.325) -- (i2b);
    \draw[rounded corners,postaction={decorate,decoration={markings,mark=at position 0.15 with {\arrow[scale=0.75]{>}},mark=at position 0.9 with {\arrow[scale=0.75]{>}}}}] (ii) |- (2.625,1.975) |- (1.875,2.375) |- (iib);
    \draw[rounded corners,postaction={decorate,decoration={markings,mark=at position 0.2 with {\arrow[scale=0.75]{>}},mark=at position 0.9 with {\arrow[scale=0.75]{>}}}}] (ii2) -| (2.425,1.775) -| (1.875,2.175) -| (ii2b);
    \node at (7,2) {\small$\hat{D}$};
    \begin{scope}[shift={(10,0)}]
      \draw (0,5*\unit) --++(0,-5*\unit) --++ (3*\unit,0) --++(0,1*\unit) node[main node,midway,label=right:{\tiny (iv)}] (iv) {}  --++(2*\unit,0) --++(0,1*\unit) --++(1*\unit,0) --++(0,2*\unit) --++(1*\unit,0);
      \node[main node,label=right:{\tiny (iii)}] (iii) at (4.5,1.75) {};
      \node[main node,label=below:{\tiny (iii)}] (iii2) at (3.375,0.75) {};
      \node[main node,label=below:{\tiny (iv)}] (iv2) at (0.625,0) {};
      \node[main node] (iiib) at (0,3.375) {};
      \node[main node] (iii2b) at (1.75,4) {};
      \node[main node] (ivb) at (0,1.75) {};
      \node[main node] (iv2b) at (0,0.375) {};
      \draw[dashed] (-0.25,2.25) --++ (5,0)  node[right=2pt] {\tiny $a_n$};
      \draw[rounded corners,postaction={decorate,decoration={markings,mark=at position 0.8 with {\arrow[scale=0.75]{>}}}}] (iv2) |- (iv2b);
      \draw[rounded corners,postaction={decorate,decoration={markings,mark=at position 0.15 with {\arrow[scale=0.75]{>}},mark=at position 0.95 with {\arrow[scale=0.75]{>}}}}] (iv) -| (1.25,1.125) -| (0.625,1.75) -- (ivb);
      \draw[rounded corners,postaction={decorate,decoration={markings,mark=at position 0.15 with {\arrow[scale=0.75]{>}},mark=at position 0.95 with {\arrow[scale=0.75]{>}}}}] (iii2) |- (2.625,1.125) |- (1.75,1.75) -- (iii2b);
      \draw[rounded corners,postaction={decorate,decoration={markings,mark=at position 0.03 with {\arrow[scale=0.75]{>}},mark=at position 0.9 with {\arrow[scale=0.75]{>}}}}] (iii) -| (4.125,2.75) -| (3.375,3.375) -- (iiib);
    \end{scope}
  \end{tikzpicture}
}

\bibliography{main}

\title{Reconstructing a bijection on the level of Le diagrams}
\date{\today}
\author{Simone Hu}
\address{Mathematical Institute \newline
  \indent University of Oxford \newline
  \indent Oxford, United Kingdom, OX2 6GG}
\email{simone.hu@maths.ox.ac.uk}

\author{Karen Yeats}
\address{Department of Combinatorics and Optimization \newline
  \indent Faculty of Mathematics, University of Waterloo \newline
  \indent Waterloo, ON, Canada, N2L 3G1}
\email{kayeats@uwaterloo.ca}

\begin{document}

\begin{abstract}
  Lukowiski, Parisi, and Williams formulated the T-duality map of string theory at a purely combinatorial level as a map on decorated permutations.  We combinatorially describe this map at the level of Le diagrams.  This perspective makes the dimension shift under the map more transparent.
\end{abstract}

\maketitle

\setcounter{tocdepth}{1}
\tableofcontents
\setcounter{tocdepth}{3}

\section{Introduction}\label{S:intro}

In this paper, we explore the combinatorics of a particular map on cells of the positive Grassmannian called {T-duality}, corresponding to the $T$-duality of string theory in the context of the positive Grassmannian and on-shell diagrams from $\mathcal{N}=4$ SYM theory.
In particular, we investigate what T-duality looks like as a map on Le diagrams, one of the combinatorial objects in bijection with positroids.

The structure of the paper is as follows.
We open with a background section, Section~\ref{S:background}, where we define T-duality at the level of decorated permutations following \cite{LPW} in Section~\ref{S:t-duality}.
In Section~\ref{S:visual} we give a combinatorial reformulation of the T-duality map directly on Le diagrams, the main result being Theorem~\ref{glueing}.  
In Section~\ref{S:proof}, we prove that the construction does indeed give the correct Le diagram, resulting in Theorem~\ref{letheorem}.
In particular, we see in Theorem~\ref{dimension} how viewing T-duality on Le diagrams directly explains the dimensional relationship between the positroid cells on either side of the map.
We briefly discuss some extensions in Section~\ref{S:discussion}.  Appendix~\ref{S:algorithm} gives another formulation of our main construction as a row-by-row algorithm.

\bpointn{Acknowledgements.}
KY is supported by an NSERC Discovery grant and by the Canada Research Chairs program.
The results of the present paper first appeared in the master's thesis of SH \cite{Hmmath}.
Both authors are grateful for the hospitality of Perimeter Institute where the writing of this work was finalized.
Research at Perimeter Institute is supported in part by the Government of Canada through the Department of Innovation, Science and Economic Development Canada and by the province of Ontario through the Ministry of Economic Development, Job Creation and Trade. 
This research was also supported in part by the Simons Foundation through the Simons Foundation Emmy Noether Fellows Program at Perimeter Institute. 

\section{Background}\label{S:background}

The Grassmannian is a classical geometric object that has been extensively studied due to its nice structure and its many connections to different areas of mathematics, including combinatorics, algebraic and differential geometry, and representation theory. 
For algebraic combinatorialists, the interest lies in the beautiful combinatorics arising from its decompositions.
Through the classic Schubert decomposition of the Grassmannian into {Schubert cells}, which can be indexed by partitions, we are led to some familiar combinatorial machinery such as Young tableaux, Schur functions, and Schubert polynomials.
A standard reference for many of these topics is~\cite{Fulton}.  
From a different decomposition, as first described by Gelfand, Goresky, MacPherson, and Serganova in~\cite{GGMS}, we are led to matroids and matroid polytopes. 
Specifically, we can divide the elements of the Grassmannian based on which Pl\"ucker coordinates are non-zero, into {matroid strata}, also known as the Gelfand-Serganova strata. 
However, unfortunately, Mn\"ev's Universality Theorem~\cite{mnev} tells us that the structure of these matroid strata can be as complicated as any algebraic variety. 

Instead of looking at the full Grassmannian, Postnikov in~\cite{tpgrass} initiated the study of a certain subset of the Grassmannian called the positive Grassmannian, by giving a combinatorial description of its cells which turned out to have a much nicer geometric structure.
This opened the door to the extensive study of the positive Grassmannian, both combinatorially and through the multitude of emerging connections with other branches of mathematics and physics. 
Like how the Grassmannian could be subdivided into its matroid strata, analogously we can do the same with the positive Grassmannian.
This gives us {positroids} and {positroid cells}, see Definition 3.2 of \cite{tpgrass}. 
In other words, we are partitioning the positive Grassmannian based on the Pl\"ucker coordinates that are strictly positive.
This is called the {positroid stratification} of the positive Grassmannian. 

For us, the interest in positroid cells lies in their rich combinatorial structure, arising from the many families of combinatorial objects that Postnikov in ~\cite{tpgrass} showed index these cells.
Beyond positroids, these objects include decorated permutations, Le diagrams, Grassmann necklaces, and equivalence classes of reduced plabic graphs. 
We will only need decorated permutations and Le diagrams.

\bpoint{Decorated permutations}

\tpointn{Definition}\label{def:decperm}
\statement{
  A \textbf{decorated permutation} of $\,[n]$ is a permutation $\pi$ on $[n]$ where every fixed-point is designated (coloured) as a \textbf{loop} (black), denoted by $\pi(i) = \underline{i}$, or a \textbf{co-loop} (white), denoted by $\pi(i) = \overline{i}$. \\
  An \textbf{anti-excedance} of $\pi$ is an element $i$ such that either $\pi^{-1}(i) > i$ or $i$ is a co-loop.
  We refer to $\pi^{-1}(i)$ as the \textbf{position} of $i$.
}

\tpointn{Example} (from~\cite{LPW})\label{ex:perm}
\statement{
  The following is a decorated permutation of $\,[8]$ in two-line notation:
  \[ \pi = \left(\begin{array}{cccccccc} 1 & 2 & 3 & 4 & 5 & 6 & 7 & 8 \\ 3 & \underline{2} & 5 & 1 & 6 & 8 & \overline{7} & 4 \end{array}\right), \]
  which has 3 anti-excedances, $1, 4, 7$, one loop, $2$, and one co-loop, $7$.
}

To give the connection between decorated permutations and cells of the positive Grassmannian, Postnikov used another object called the {Grassmann necklace}, which can be read off from the bases of the positroid, see \textsection 16 of~\cite{tpgrass}. For decorated permutations, the result is as follows.

\tpointn{Theorem} (from Lemma 16.2, Theorem 17.1 of~\cite{tpgrass})
\statement{
  Decorated permutations of $\,[n]$ with $k$ anti-excedances index the cells of the $(k,n)$ positive Grassmannian, $\:\Gr_{k,n}^{\geq 0}$, and we denote by $S_{\pi}$ the positroid cell indexed by $\pi$.
}

Going back to Example~\ref{ex:perm}, this means that $\pi$ indexes the cell $S_{\pi}$ of $\:\Gr_{3,8}^{\geq 0}$.

\bpoint{Le diagrams and pipe dreams}\label{SS:le-def}

Decorated permutations are simple and succinct objects that encode positroids, but one property that they do not easily see is the dimension of the associated positroid cell.
For this, we need the next family of combinatorial objects called Le diagrams. 

\tpointn{Definition} (Definition 6.1 of~\cite{tpgrass})\label{def:le}
\statement{
  A \,\textbf{\BLe-diagram} (or Le diagram), is a filling $D$ of a Young diagram of shape $\lambda$ with $\,0$'s and $\leplus$'s such that $D$ avoids the \textbf{\BLe-configuration}:
  \[ \leconfig \] 
  That is, no $\,0$ has both a $\leplus$ above and to the left of it, which we refer to as the \textbf{\BLe-condition}.\\
  For $0 \leq k \leq n$ we say that the \Le-diagram $D$ is of \textbf{type (k,n)} if the shape $\lambda$ fits inside a $k \times (n-k)$ rectangle.
}

An example of a \Le-diagram $D$ is given in Figure~\ref{fig:ex-le}.

In \textsection 20 of~\cite{tpgrass}, Postnikov gave two bijections between \Le-diagrams and decorated permutations, the first through associating a series of other objects (hook diagrams, networks, and plabic graphs) to \Le-diagrams.
The second, which we describe here, uses an algorithm from \textsection 19 of~\cite{tpgrass} going through a slightly different object called pipe dreams.

Given a \Le-diagram $D$ of type $(k,n)$, we associate a decorated permutation $\pi_D$ on $[n]$ as follows:
\begin{enumerate}
  \item In $D$, we replace each $0$ with a cross $\,\smallcross\,$ and each $\leplus$ with an elbow joint $\,\smallelbow\;$:
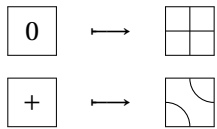
    \begin{align*}
      \lebox{0}\quad&\longmapsto\quad \lebox{\cross{}} \\[4pt]
      \lebox{+} \quad&\longmapsto\quad \lebox{\elbow{}}
    \end{align*}
  \item View the south-east (SE) border of $D$ as a lattice path with $n$ steps, and label the edges with $1, \ldots, n$ along this path from the top-right corner of the bounding $k \times (n-k)$ rectangle to the bottom-left corner of the bounding rectangle.  
  \item Then label the $n$ edges of the north-west (NW) border of $D$ so that, viewing $D$ as a grid, the rows and columns have the same labels (the opposite horizontal/vertical edges are labelled the same).
    We call this diagram $P$ the \textbf{pipe dream} associated to $D$.
  \item To get the decorated permutation $\pi$ associated to $P$ and $D$, we follow the "pipes" of $P$ from the SE border to the NW border. That is, $\pi(i) = j$ if the pipe starts at $i$ and ends at $j$.
    A horizontal pipe starting and ending at $i$, where $i$ labels vertical edges, is denoted a co-loop $\pi(i) = \overline{i}$.
    A vertical pipe starting and ending at $i$, where $i$ labels horizontal edges, is denoted a loop $\pi(i) = \underline{i}$.
\end{enumerate}

To illustrate this algorithm, we use an example.
\tpointn{Example}\label{ex:le}
\statement{
  Given the \Le-diagram $D$ of type $(3,8)$ in Figure~\ref{fig:ex-le}, we label the SE and NW borders with $1, \ldots, 8$ as above, and replace the $0$'s and $\leplus$'s with crosses and elbow-joints.
  This gives the pipe dream $P$ in Figure~\ref{fig:ex-pipe}, where arrows are added to show the direction to follow the pipes for the permutation. \\
  \begin{figure}[ht]
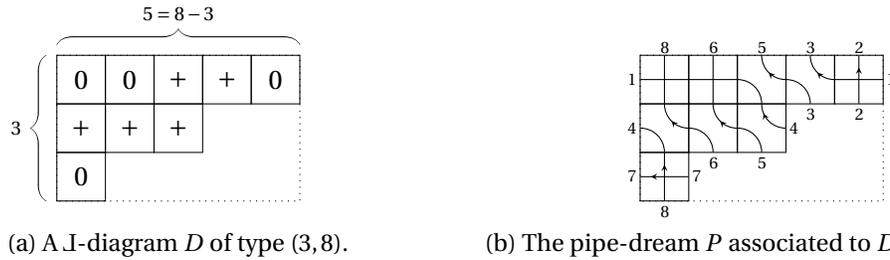

  \centering
  \begin{subfigure}[b]{.48\linewidth}
    \centering
    \[ \exle \]
    \caption{A \Le-diagram $D$ of type $(3,8)$.}
    \label{fig:ex-le}
  \end{subfigure}
  \begin{subfigure}[b]{.48\linewidth}
    \centering
    \[ \expipe \]
    \caption{The pipe-dream $P$ associated to $D$ in Figure~\ref{fig:ex-le}.}
    \label{fig:ex-pipe}
  \end{subfigure}
  \caption[The \Le-diagram and pipe dream associated to Example~\ref{ex:perm}.]{The \Le-diagram and pipe dream associated to Example~\ref{ex:perm}. The arrows on the pipes indicate the direction to follow to read off the decorated permutation.}
  \label{fig:ex-le-pipe}
\end{figure} \\
  To read off the decorated permutation, we follow the pipes in the direction of the arrows to get
  \[ \pi_D = \left(\begin{array}{cccccccc} 1 & 2 & 3 & 4 & 5 & 6 & 7 & 8 \\ 3 & \underline{2} & 5 & 1 & 6 & 8 & \overline{7} & 4 \end{array}\right), \]
  which is the decorated permutation from Example~\ref{ex:perm}.
}

Observe that rows of all $0$'s in $D$ correspond to co-loops in $\pi_D$, while columns of all $0$'s correspond to loops.  Pipes only ever move up or to the left, and avoiding the \Le-configuration in $D$ corresponds to pipes in $P$ crossing at most once, and furthermore, once two pipes cross they never subsequently share a box even without crossing (via an elbow joint).

It turns out that this map is indeed a bijection. One of the reasons to look at \Le-diagrams is because of how easily they display the dimension of the positroid cell that they index.
In particular, the dimension of the corresponding positroid cell is determined by counting the number of $\leplus$'s in the \Le-diagram.

\tpointn{Theorem} (Theorem 6.5, Corollary 20.1, Theorem 20.3 of~\cite{tpgrass})
\statement{
  The map $D \mapsto \pi_D$ is a bijection from \Le-diagrams of type $(k,n)$ to decorated permutations of $[n]$ with $k$ anti-excedances.
  Therefore, \Le-diagrams of type $(k,n)$ index the cells of the $(k,n)$ positive Grassmannian, $\:\Gr_{k,n}^{\geq 0}$.
  Furthermore, let $S_D$ be the positroid cell indexed by $D$.
  Then the dimension of the cell $S_D$ is the number of $\:\leplus$'s in $D$.
}

Going back to Example~\ref{ex:le}, this means that $D$ indexes the cell $S_{D}$ of $\:\Gr_{3,8}^{\geq 0}$, which corresponds to the cell $S_{\pi}$ from Example~\ref{ex:perm}, and the dimension of $S_D$ is $5$ since there are 5 $\leplus$'s in $D$. \\

\bpoint{T-duality on decorated permuations}\label{S:t-duality}

The positive Grassmannian has deep connections to many different areas of mathematics and physics. 
We have already mentioned some of the mathematical connections, others include oriented matroids, polytopes and polyhedral subdivisions, non-crossing partitions, lattice paths, cluster algebras and quantum algebras. The positive Grassmannian can also be studied in the spirit of Schubert calculus, through varieties, flags, and symmetric functions.
On the physics side, there have been applications to KP solitons, types of asymmetric exclusion processes, and scattering amplitudes (via the amplituhedron and Wilson loop diagrams).
We refer the reader to ~\cite{tpgrass, post, kps, grassampl, ampl, asep, LPW, PSBW, agarwala_fryer_yeats_2022_I, agarwala_fryer_yeats_2022_II} and the references therein for further details.  

The aspect of the physics we are interested in is how T-duality from string theory, via the connection with the amplituhedron, then manifests itself combinatorially.
Lukowski, Parisi, and Williams~\cite{LPW} showed that at the level of decorated permutations, T-duality becomes the following very elegant map.

\tpointn{Definition} (Definition 5.1 of~\cite{LPW})\label{def:t-duality}
\statement{
  The \textbf{T-duality} map from loopless decorated permutations on $[n]$ to co-loopless decorated permutations on $[n]$ is defined as
  \begin{align*}
    \pi \quad&\longmapsto\quad \hat\pi \\
    (a_1, a_2, \ldots, a_n) \quad&\longmapsto\quad (a_n, a_1, \ldots, a_{n-1})
  \end{align*}
  where the permutations are written in one-line notation, and any fixed points in $\hat\pi$ are declared to be loops.
  That is, for a given loopless $\pi$, we have $\hat\pi(i) = \pi(i-1)$ where all fixed points are loops, and we call $\hat\pi$ the \textbf{T-dual} decorated permutation.
}

\tpointn{Lemma} (Lemma 5.2, Proposition 5.17 of~\cite{LPW})
\statement{
  The T-duality map $\pi \mapsto \hat\pi$ is a bijection between loopless decorated permutations on $[n]$ with $k+1$ anti-excedances and co-loopless decorated permutations on $[n]$ with $k$ anti-excedances. \\\\
  Equivalently, the T-duality map is a bjiection between loopless positroid cells $S_{\pi}$ of $\:\Gr_{k+1,n}^{\geq 0}$ and co-loopless positroid cells $S_{\hat\pi}$ of $\:\Gr_{k,n}^{\geq 0}$.
  Furthermore, we have the following dimensional relationship
  \[ \dim(S_{\hat\pi}) - 2k = \dim(S_{\pi}) - (n-1), \]
  where in particular, if $\dim(S_{\pi}) = n-1$ then $\dim(S_{\hat\pi}) = 2k$.
}

While this dimensional relationship was proven to exist between T-dual loopless cells of $\Gr_{k+1,n}^{\geq 0}$ and co-loopless cells of $\Gr_{k,n}^{\geq 0}$, it was not clear where it was coming from.
The desire to better understand this relationship motivated the question of viewing T-duality as a map on \Le-diagrams, in which one very easily sees the dimension of the positroid cells they index.

The key result of \cite{LPW} (Theorem 6.5) is that the T-duality map provides a bijection between BCFW tilings of a particular hypersimplex, $\:\Delta_{k+1,n}$, and BCFW tilings of the particular amplituhedron $\mathcal{A}_{n,k,2}$.
After the original appearance of our work in \cite{Hmmath}, Parisi, Sherman-Bennett, and Williams~\cite{PSBW} extended the work done in~\cite{LPW} and proved the main conjecture in~\cite{LPW} which strengthens the bijection to all positroid tilings.
Their proof utilized looking at the T-duality map via plabic graphs and plabic tilings.
Together with our result, then, the T-duality map has a direct combinatorial formulation for decorated permutations, Le diagrams and plabic graphs. See Section~\ref{S:discussion} for further discussion.

\section{T-duality on the level of Le diagrams}\label{S:visual}

The direction of T-duality we will look at is $\hat\pi \mapsto \pi$, where $\hat\pi$ is a co-loopless permutation on $[n]$ with $k$ anti-excedances and $\pi$ is a loopless permutation on $[n]$ with $k+1$ anti-excedances.
On \Le-diagrams, we are, thus, looking at the map from $\hat{D} \mapsto D$, where $\hat{D}$ is the \Le-diagram associated to $\hat\pi$ and $D$ is the \Le-diagram associated to $\pi$.

We will define our map by showing how to take each column of $\hat{D}$ that contains at least one $\leplus$ and convert it into an explicit configuration of $\leplus$'s which when glued together characterize $D$ (see Definition~\ref{glueing}).  This gives an explicit, combinatorial form for the T-duality map on \Le-diagrams.  In Appendix~\ref{S:algorithm} we give an alternate description of this map which acts row-by-row and is more algorithmic. 

\bpoint{Notation}

Throughout the rest of this paper, we will refer to the rows and columns of a \Le-diagram $D$ by the same labelling as that which gives the associated decorated permutation (i.e from its pipe dream).
Boxes in $D$, and the $k \times (n-k)$ rectangle, will then be referred to by their coordinates $(i,j)$ under this labelling.
A box is considered as "existing" if it is a valid box to be filled within the shape $\lambda$.
Note that all such valid boxes have $i < j$.
An example of this notation is given in Example~\ref{ex:notation}.
Finally, since we are dealing with two \Le-diagrams, one on each side of the map, we refer to the "corresponding" row/column as the row/column with the same label on the opposite side of the map.

We will also order boxes in columns from top to bottom and in rows from right to left, in accordance with labels going from smallest to largest.
For rows/columns, having $i < j$ would mean $i$ is to the right and/or above $j$.
For boxes, "first" refers to the right/top-most in a row/column, "last" refers to the left/bottom-most in a row/column and "next" refers to the next box to the left/below in a row/column.

\tpointn{Example}\label{ex:notation}
\statement{
  Going back to Example~\ref{ex:le}, we have the following labelling of the \Le-diagram $D$
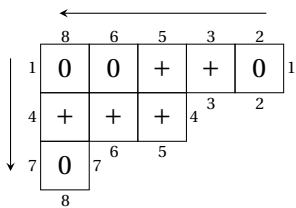
  \[ \exnotation \]
  where there are $\leplus$'s in boxes $(1,3)$, $(1,5)$, $(4,5)$, $(4,6)$, and $(4,8)$.
  Notice box $(4,3)$ does not exist since it is outside of the shape $\lambda$ of $D$.
  By the nature of the labelling, we also never have boxes of the form $(i,i)$.
  The arrows indicate the direction of the ordering of boxes, going from first to next to last.
}

\bpoint{Preliminaries and shape of \texorpdfstring{$D$}{D}}\label{SS:shape}

We start with a co-loopless decorated permutation $\hat\pi$ on $[n]$ with $k$ anti-excedances, which we denote in two-line notation as
\[\hat\pi = \left(\begin{array}{cccc} 1 & 2 & \cdots & n \\ a_n & a_1 & \cdots & a_{n-1}\end{array}\right). \]
Let $\{b_1, \ldots, b_k\}$ be the $k$ anti-excedances of $\hat\pi$ ordered such that
$b_1 < \cdots < b_k$ (recall that $b_u$ is an anti-excedance if $\hat\pi^{-1}(b_u) > b_u$, where we can't have $\hat\pi(b_u) = \bar{b}_u$ since $\hat\pi$ is co-loopless).
In particular, let $i_u = \hat\pi^{-1}(b_u)$ be the position of $b_u$.
Then we have $b_u = \hat\pi(i_u) = a_{i_u -1}$ for $1 \leq u \leq k$.

\begin{figure}[th]
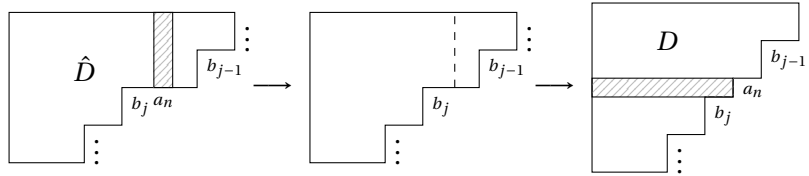

  \centering
  \[ \constructD \]
  \caption[Steps to construct the shape of $D$ from $\hat{D}$.]{Steps to construct the shape of $D$ from $\hat{D}$. Remove column $a_n$ from $\hat{D}$ and insert a row labelled $a_n$ where the dashed line is, making sure that the new boundary path is in the correct order. Here $j$ is the index such that $b_{j-1} < a_n < b_j$.}
  \label{fig:construct-D}
\end{figure}

We denote the associated $\Le$-diagram by $\hat{D}$ with shape $\hat\lambda$.
Since $\hat\pi$ is co-loopless, we have that every row in $\hat{D}$ has $\geq 1 \;\leplus$'s.
In particular,
\[ \hat\lambda = \left(\hat\lambda_1, \cdots, \hat\lambda_k\right), \quad\text{where }\hat\lambda_u = (n-k) - (b_{u} - u) \text{ for } 1 \leq u \leq k \]
and every $\hat\lambda_u \geq 1$.
We also have that the rows of $\hat{D}$ are thus labelled by these $b_{u}$, as in the left diagram of Figure~\ref{fig:construct-D}. Note that there are no rows of size 0 as $\hat\pi$ is co-loopless.

We want to produce a $\Le$-diagram $D$ of type $(k+1, n)$ and dimension ${\dim(S_{\hat{D}}) -2k + (n-1)}$ with associated decorated permutation $\pi$ of $[n]$ given by
\[\pi = \left(\begin{array}{cccc} 1 & 2 & \cdots & n \\ a_1 & a_2 & \cdots & a_{n}\end{array}\right), \]
which has $k+1$ anti-excedances and is loopless.
The loopless condition means that $D$ has to have $\geq 1 \;\leplus$'s in each of the $n-(k+1)$ columns.

To determine the shape $\lambda$ of $D$, which only depends on the anti-excedances of $\pi$, consider the following:
\begin{itemize}
  \item $a_n$ is not an anti-excedance of $\hat\pi$ since $\hat\pi$ is co-loopless, $\hat\pi(1) = a_n$ and $1 \not> a_n$.
    In particular, $a_n$ is the label of a column in $\hat{D}$.
  \item Under T-duality, $b_u$ for $1 \leq u \leq k$ stays an anti-excedance as $i_u > b_u = \hat\pi(i_u) = \pi(i_u-1)$.
    Since $\pi$ is loopless, we have either $\pi(i_u - 1) < i_u-1$ or $\pi(i_u-1) = i_u-1$ where $i_u-1$ must be a co-loop.
    In either case, $b_u$ is an anti-excedance of $\pi$.
  \item $a_n$ is always an anti-excedance of $\pi$ since $\pi$ is loopless.
  \item There are no other anti-excedances of $\pi$ since $\pi(i) = \hat\pi(i+1) \geq i+1 > i$ for all $1 \leq i < n$ with $i+1 \not\in \{i_1, \ldots, i_k\}$.
\end{itemize}
Based on these observations, we have that $\{b_1, \ldots, b_k\} \cup \{a_n\}$ are the $k+1$ anti-excedances of $\pi$.
In particular, the labels of the rows of $D$ (including rows of length 0) are thus exactly the same as $\hat{D}$, with the addition of $a_n$.
Then, the shape of $D$ can be constructed from $\hat{D}$ by removing the column labelled $a_n$ and inserting in a row labelled $a_n$ in the appropriate position, maintaining the order of the labels of the new boundary lattice path, see Figure~\ref{fig:construct-D}.

The shape $\lambda$ of $D$ is then (including 0 sized parts)
\[ \lambda = \left(\lambda_1, \cdots, \lambda_{k+1}\right) \quad\text{where}\quad
  \lambda_{u} = \begin{cases} \hat\lambda_u - 1 & 1 \leq u \leq j-1, \\
  n - (k+1) - (a_n - j) & u = j,\\
  \lambda_{u} = \hat\lambda_{u-1} & j+1 \leq u \leq k+1, \end{cases} \]
and $j$ is the index such that $b_{j-1} < a_n < b_j$.
If $a_n > b_u$ for all $1 \leq u \leq k$, then let $j = k+1$.
We have $0 \leq \lambda_u \leq n - (k+1)$ for all $1 \leq u \leq k+1$, and at most $k+1$ non-zero parts, as needed.
Thus the order of rows (and anti-excedances) of $D$ is $\{b_1, \ldots, b_{j-1}, a_n, b_j, \ldots, b_k\}$, as in the rightmost diagram of Figure~\ref{fig:construct-D}.  When $a_n = n$, we would have a row of size 0 (a co-loop) labelled by $n$.
Note that we always have either $b_1 = 1$ or $a_n = 1$, and thus the first row of $D$ will be labelled with $1$ and is a full row of size $\lambda_1 = n - (k+1)$. 

\bpoint{Building blocks}

Our construction of $D$ is built of 2 distinct shapes; what we will call $\mathbb{L}$-shapes and strings of $\leplus$'s.
To create the $\mathbb{L}$-shapes, we will look at columns $\ell \neq a_n$ in $\hat{D}$ with at least one $\leplus$.
We refer to $\leplus$'s in $\hat{D}$ that are not a leftmost (last) $\leplus$ in its row as \textbf{non-last}.

\tpointn{Definition}\label{shapes}
\statement{
  Let $\ell \neq a_n$ be a column in $\hat{D}$ with at least one $\leplus$. 
  Let $(f, \ell)$ be the first (topmost) $\leplus$ in this column and $(g, \ell)$ be the last (bottom-most).
  To each such $\ell$, we place in $D$ a corresponding shape
    \[ \lshapelabels \]
  which we call an \textbf{$\;\mathbb{L}$-shape}, specified as follows.\\
  For the indices:
  \begin{itemize}
        \item Let $m$, possibly equal to $\ell$, be the right-most column such that all columns strictly between $\ell$ and $m-1$ in $\hat{D}$ are of the same height as column $\ell$ and are filled with only 0's.
        \item Let $b_B = \max\{g, a_n\}$ if $a_n<\ell$, otherwise let $b_B = g$.
        \item Let $b_T= f$ if $f<a_n < \ell$.  Otherwise, let $h$ be the first row above $f$ with a $\leplus$ to the left of column $\ell$ in $\hat{D}$, if such a row exists, and then let $b_T=\max\{a_n, h\}$ when $a_n <f$ and let $b_T = h$ when $a_n>\ell$.  Note that $h$ always exists when $a_n>\ell$ as $\hat{D}$ has a $\leplus$ at $(1,a_n)$. 
  \end{itemize}
  For the $\leplus$'s:
  \begin{itemize}
    \item For the horizontal part: fill boxes $(b_B, \ell -1), \ldots, (b_B, m)$ with $\leplus$'s (if $m = \ell$ there are no such boxes).
    \item For the vertical part: 
    \begin{itemize}
        \item fill box $(a_n, \ell)$ with $\leplus$ if $b_T \leq a_n \leq b_B$ or fill box $(b_T, \ell)$ with $\leplus$ otherwise (i.e. $b_T \neq f, a_n$) 
        \item fill box $(b, \ell)$ with $\leplus$ if there is a non-last $\leplus$ at $(b, \ell)$ in $\hat{D}$, or if there is a last $\leplus$ but $b$ is a row with a $\leplus$ in column $a_n$, for rows $b$ from $f$ to $g$ in $\hat{D}$.
    \end{itemize}       
  \end{itemize}
  Fill all other boxes of the $\mathbb{L}$-shape with $0$s.
}

\tpointn{Remark}\label{rem L rems}
\statement{%
  \begin{itemize}
    \item There is always a $\leplus$ in box $(b_T, \ell)$ in $D$.
    \item We always have $b_T \leq f \leq g \leq b_B$ with $b_T \neq b_B$ and thus $\mathbb{L}$-shapes are well-defined. 
    \item The horizontal string of $\leplus$'s corresponds to consecutive columns of all $0$'s in $\hat{D}$ to the right of any column with at least one $\leplus$.
  \end{itemize}
}

Notice that these $\mathbb{L}$-shapes cover those columns of $D$ which correspond to columns of $\hat{D}$ (ignoring $a_n$) with at least one $\leplus$ as well as columns of all $0$'s to the right of these and of the same height. 
What's left are the columns of all $0$'s which are to the left of any columns of the same height that contain $\leplus$'s.
We cover these columns by strings of $\leplus$'s.

\tpointn{Definition}\label{strings}
\statement{
  For consecutive columns of all 0's in $\hat{D}$ of the same height and preceding any column of the same height containing a $\leplus$, we will place in $D$ a corresponding \textbf{string of $\leplus$'s}
    \[ \stringplus \]
  and these will be the only $\leplus$'s in these columns.
}

We leave the specification of which row these strings of $\leplus$'s will go to Definition~\ref{glueing}, where it becomes clearer how the two shapes associated to columns in $\hat{D}$ glue together to form $D$.

These strings of $\leplus$'s behave like the horizontal part of an $\mathbb{L}$-shape.  With appropriate conventions for the indices of Definition~\ref{shapes} we can view them as degenerate $\mathbb{L}$-shapes without the vertical part, but this becomes more intricate than simply considering them separately.

To give some intuition about what these shapes are doing, the idea here is that:
  \begin{itemize}
    \item The vertical part of an $\mathbb{L}$-shape has almost the same configuration of $\leplus$'s as the $\leplus$'s in $\hat{D}$ in that column $\ell$ with two exceptions.  The first exception is that the vertical part of the $\mathbb{L}$-shape is expanded to either include the new row $a_n$ or to a row directly above whose last $\leplus$ has not yet passed, and adding a $\leplus$ in that new row. The other difference is we instead place $0$'s in the $\mathbb{L}$ for last $\leplus$'s in column $\ell$ in $\hat{D}$ in rows without a $\leplus$ in column $a_n$.
    \item Then the horizontal part of $\mathbb{L}$-shapes start from a column with at least one $\leplus$ in $\hat{D}$ and extend to the right with a string of $\leplus$'s to the previous column of $\leplus$'s in $\hat{D}$ or the start of a row. 
    \item Strings of $\leplus$'s not in the $\mathbb{L}$-shapes extend to the left until the start of the next row below, and to the right until a column from $\hat{D}$ with at least one $\leplus$.
  \end{itemize}

From Definitions~\ref{shapes} and~\ref{strings}, we immediately get the following statement.

\tpointn{Proposition}\label{number}
\statement{
  There is at least one $\leplus$ in every column of $D$.
  More specifically,
  \begin{enumerate}
    \item \label{plusi} there is exactly one $\leplus$ in columns corresponding to those in $\hat{D}$ with no $\leplus$'s, and
    \item \label{plusii} there are $s_{\ell} + t_{\ell} + 1$ $\leplus$'s in columns $\ell$ corresponding to those in $\hat{D}$ with at least one $\leplus$, where in column $\ell$ in $\hat{D}$ there are $s_{\ell}$ non-last $\leplus$'s and $t_{\ell}$ last $\leplus$'s in rows with a $\leplus$ in column $a_n$.  
  \end{enumerate}
}

Now that we have our building blocks, what's left is to piece them together.

\bpoint{Gluing these shapes together}

How these shapes are glued together is even nicer than one might first guess from the indices in Definition~\ref{shapes}.  We will use this to define in which rows the strings of $\leplus$'s from Definition~\ref{strings} appear and prove a nice characterization of how the $\mathbb{L}$-shapes glue.  First we need a definition.

\tpointn{Definition}\label{sections}
\statement[eq]{
  Given a Young diagram $D$ of shape $\lambda = (\lambda_1, \cdots, \lambda_k)$, we partition $D$ into $k$ rectangles called \textbf{sections} where each section has dimension $j \times (\lambda_j - \lambda_{j+1})$ for $1 \leq j \leq k$, and is bounded by two rows (one possibly empty). We let $\lambda_{k+1} = 0$.
    \[ \sectionrow \]
  We name each section by its last row and include empty sections in the count of $k$ (when some $\lambda_j = \lambda_{j+1}$).
}

\tpointn{Definition/Theorem}\label{glueing}
\statement{
  $D$ is made up of gluing together the $\;\mathbb{L}$-shapes of Definition~\ref{shapes} and the strings of $\leplus$'s of Definition~\ref{strings} where in each (non-empty) section $b$, we have a chain of shapes in the following form:
    \[ \chain \]
    where the bottom right of each shape is glued to the last $\leplus$ in the vertical part of the previous shape, except the first which is glued to row $b$.
  More precisely, we have the following properties for each section:
  \begin{itemize}
  \item There are either none or exactly one string of $\leplus$'s in the chain, and if there is one it is leftmost in the chain.  The other shapes in the chain are all $\;\mathbb{L}$-shapes. There may be $0$ $\;\mathbb{L}$-shapes, but there must be at least one shape total in the chain.
    \item $\mathbb{L}$-shapes are glued on the left to other $\;\mathbb{L}$-shapes at the last $\leplus$ in the vertical part of a previous $\;\mathbb{L}$, except the first $\;\mathbb{L}$ in the chain which is glued to row $b$.
    \item The string of $\leplus$'s is either glued on the left to the last $\;\mathbb{L}$-shape in the chain at the last $\leplus$ in the vertical part, or glued to row $b$, in which case it fills the whole width of a section. This defines the vertical location of the string of $\leplus$'s.
  \end{itemize}
  The rest of the boxes in $D$ (the shaded regions) are filled with $0$'s.
}

\begin{proof}
  Putting together Definitions~\ref{shapes} and~\ref{strings}, $D$ is made up of the following two types of blocks, one consisting of $\mathbb{L}$-shapes where $\ell$ corresponds to a column with at least one $\leplus$ in $\hat{D}$, and the other consists of strings of $\leplus$'s which corresponds to consecutive columns of all 0's in $\hat{D}$.\\
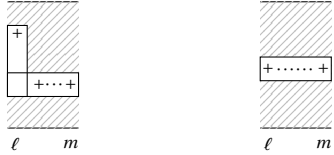
  \[ \blocks \]
  In the blocks, the shaded regions represent boxes filled with all 0's, which extends to fill the rest of the rows in columns $m$ to $\ell$ in the shape $\lambda$.
  The upper and lower lines represent the border of $\lambda$.
  As every column in $D$ was originally a column in $\hat{D}$ and the new row $a_n$ is accounted for in the $\mathbb{L}$-shapes, these blocks fit in and cover each column of the shape of $D$. 

  Thus each section of $D$ is made up of a non-zero chain of $\;\mathbb{L}$-shapes and potentially one string of $\leplus$'s.  Observe the following:
  \begin{itemize}
    \item If there is a column with at least one $\leplus$ in section $b$ of $\hat{D}$, then there will be a first (right-most) $\mathbb{L}$ in section $b$ of $D$ which has $m-1=b_B =b$:
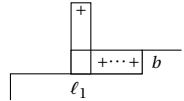
        \[ \firstl \]
      Here $\ell_1$ is the first column of $\leplus$'s in $\hat{D}$ in section $b$.
    \item When section $b$ in $\hat{D}$ consists of columns of all $0$'s, then the string of $\leplus$'s extends to fill the whole bottom row $b$ of the section:
        \[ \sectiononlystring \]
    \item If the column directly before the next row below $b$ has at least one $\leplus$ in $\hat{D}$, then there is no string of $\leplus$'s in section $b$. 
      Otherwise, there is exactly one string of $\leplus$'s in section $b$.
  \end{itemize}

  What we need to show is how the shapes glue together.
  Consider a particular section $b$.
  First, we already saw the special case when a section contains only columns of all $0$'s, we can assume there is at least one $\mathbb{L}$ in section $b$.
  We also saw that the first $\mathbb{L}$ in the section will always glue to the beginning of the row.
  The desired gluing rule is the definition for the gluing of the string of $\leplus$'s, so all we need to show is how $\mathbb{L}$'s glue on the left to other $\mathbb{L}$'s.

  Consider an $\mathbb{L}$-shape with vertical part in column $\ell$ (or with no vertical part) and has right-most and bottom-most box at $(b_B, m)$.
  We want to see where $b_B$ is in relation to the previous $\mathbb{L}$, with vertical part in column $m-1$ and say in rows $i$ to $j$:
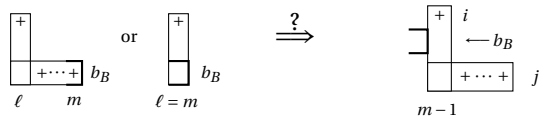
  \[ \glueproofone \]
  By construction of $b_B$, either $b_B = a_n$ or $b_B$ is a row such that in $\hat{D}$ there is a $\leplus$ in box $(b_B, \ell)$ with all $0$'s below it in column $\ell$. 
  
  In the latter case, by the \Le-condition on $\hat{D}$, not only are there only $0$'s below $(b_B, \ell)$, but all boxes to the left of those $0$'s must also be $0$'s.  Therefore, if we look at column $m-1$ in $\hat{D}$, any $\leplus$'s below row $b_B$ must be the last $\leplus$'s of their row.
  Additionally, these rows all do not have a $\leplus$ in column $a_n$ as by definition of $b_B$ we have either $a_n < b_B < m-1$, or $a_n > \ell > m-1$ but in the latter case as we already observed there are no $\leplus$'s after column $m-1$. 
  Consequently, the $\mathbb{L}$-shape for column $m-1$ will have all $0$'s in its vertical part below row $b_B$.
  Now as $\hat{D}$ has a $\leplus$ in box $(b_B, \ell)$ and since $\hat{D}$ is a \Le-diagram, either there is a non-last $\leplus$ at $(b_B, m-1)$, in which case the $\mathbb{L}$-shape for column $m-1$ also has a $\leplus$ there, or the first $\leplus$ in column $m-1$ of $\hat{D}$ is strictly below $b_B$ say in row $f$, in which case the $\mathbb{L}$-shape for column $m-1$ will have row $i = b_B$ (by the condition on $h$ in Definition~\ref{shapes} for column $m-1$). 
  
  If $b_B = a_n$, by construction there must be a row $g < a_n$ such that in $\hat{D}$ there is a $\leplus$ at box $(g, \ell)$ with all $0$'s below and, as before, all boxes to the left of those $0$'s must also be $0$'s. 
  By a similar analysis, the $\mathbb{L}$-shape for column $m-1$ will have all $0$'s in its vertical part for rows below $a_n$. 
  Now in $\hat{D}$ column $m-1$, if the first $\leplus$ occurs in a row $> a_n$ or if the last $\leplus$ occurs in a row $< a_n$ then by construction the $\mathbb{L}$-shape for column $m-1$ will have $i=a_n$ or $j=a_n$ respectively. 
  Otherwise, we have that $i < a_n < j$. 
  Regardless, the $\mathbb{L}$-shape for column $m-1$ will thus have a $\leplus$ in box $(a_n, m-1)$. 
  
  In both cases we get the following relation (where $j$ could be equal to $b_B$):
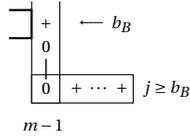
  \[ \glueprooftwo \]
  Thus we have that $\mathbb{L}$-shapes, and therefore also strings of $\leplus$'s, glue to the left of $\mathbb{L}$-shapes at the last (bottom-most) $\leplus$ in the vertical part of the $\mathbb{L}$.
\end{proof}

\bpoint{An example}\label{ex:algorithm}

\begin{figure}[h]
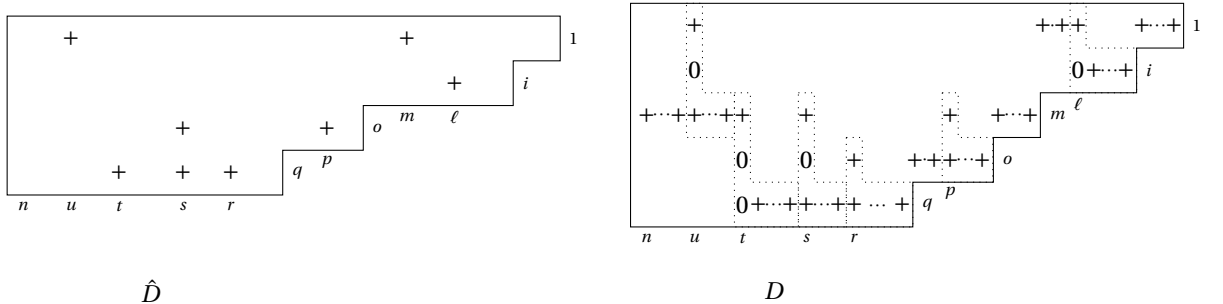

  \centering
  \[ \algexDshapes \]
  \caption[An example of ${D}$ vs. $\hat{D}$.]
  {An example of $\hat{D}$ vs. ${D}$. 
  On the left, we have a \Le-diagram $\hat{D}$ of type $(4,n)$ and dimension $2k = 8$, where all unmarked boxes are filled with $0$'s.
  On the right, we have our constructed $D$, where the dotted lines are outlining the $\mathbb{L}$-shapes associated to each column of $\leplus$'s in $\hat{D}$, except column $a_n = m$.}
  \label{fig:algex}
\end{figure}

We illustrate this construction through an example shown in Figure~\ref{fig:algex}. 
\\
To briefly check that this $D$ is indeed the right diagram for the given $\hat{D}$:
\begin{itemize}
  \item $D$ is a \Le-diagram of type $(5,n) = (k+1,n)$ (notice it avoids the \Le-configuration and has $5$ rows).
  \item Every column has at exactly one $\leplus$ except columns $p,r,s,u$, which have two $\leplus$'s.
    Thus $D$ is loopless and of dimension 
    $ \dim(S_D) = (n-5) + 4 = n - (k+1) + k = n-1$ (recall this is the number of $\leplus$'s), 
    as there are $n-5$ columns and 4 columns with an extra $\leplus$.
    This gives the correct relation since we wanted $\dim{S_{\hat{D}}} - 2k = \dim(S_D) - (n-1)$, recalling that $\dim(S_{\hat{D}}) = 8 = 2k$.
  \item One can check that the associated decorated permutation $\pi$ to $D$, in two-line notation, is
    \[\indent\indent\left(\arraycolsep=1pt\begin{array}{*{23}{c}} 1 & {\cdots} & i-1 & {\cdots} & \ell-1 & {\cdots} & m-1 & {\cdots} & o-1 & {\cdots} & p-1 & {\cdots} & q-1 & {\cdots} & r-1 & {\cdots} & s-1 & {\cdots} & t-1 & {\cdots} & u-1 & {\cdots} & n  \\ 2 & & \ell & & \circled{i} & & u & & p & & s & & r & & \circled{o} & & t & & \circled{q} & & \circled{1} & & \circled{m} \end{array}\right), \]
      where the $\;\cdots\;$ denotes $\pi(a) = a+1$ for all $a$ in-between the explicitly written values.
      Notice $\pi$ is loopless and has $5$ anti-excedances (circled).
      Shifting the bottom-line one to the right (and wrapping around) exactly gives $\hat\pi$, the decorated permutation associated to $\hat{D}$.
\end{itemize}

\bpoint{\texorpdfstring{$D$}{D} is indeed a \texorpdfstring{\BLe-}{Le }diagram}

To begin our proofs that this construction is well defined and agrees with T-duality, we first verify that under this filling $D$ avoids the \Le-configuration and thus is a valid \Le-diagram.

\tpointn{Theorem}\label{Dlediag}
\statement[eq]{
  Under this filling, $D$ is a \Le-diagram of type $(k+1, n)$.
}

\begin{proof}
  By the construction of the shape of $D$ in Section~\ref{SS:shape}, we have that $D$ is of type $(k+1,n)$. 
  We just need to show that there cannot be a \Le-configuration in $D$. 
  Suppose in $D$ we have the following
  \[ \leconfiglabels \]
  where the boxes indicated by the dots are filled with $0$'s.

  As Theorem~\ref{glueing} tells us that $\mathbb{L}$-shapes and strings of $\leplus$'s are glued together such that the horizontal part of a shape is glued to the last $\leplus$ in the vertical part of an $\mathbb{L}$ (or to a row), without loss of generality we only need to consider when both $\leplus$'s are in vertical parts of $\mathbb{L}$-shapes.  
  
  By the construction of the $\mathbb{L}$-shape for column $m$, there are two cases
  \begin{enumerate}
    \item either $j= a_n$
    \item \label{case 2} or $j\neq a_n$ in $\hat{D}$ there must be a $\leplus$ in box $(j, o)$ for some column $o \geq m > \ell$.
  \end{enumerate}
  First suppose $j = a_n$. Since there is no $\leplus$ at $(a_n, \ell)$ in the $\mathbb{L}$-shape for column $\ell$, we must be in the case that $a_n$ is not between the $b_T$ and $b_B$ for that $\mathbb{L}$-shape and since there is a $\leplus$ at $(i, \ell)$ this means that the whole $\mathbb{L}$-shape for column $\ell$ must be above $a_n$. In particular, in $\hat{D}$ the last $\leplus$ in column $\ell$ must be in a row $i$. Consequently, by the choice of $b_B$ for the $\mathbb{L}$-shape at column $\ell$, we must have $\ell < a_n$. 
  But this is impossible as then there could not have been a box at $(a_n, \ell)$ in $D$.
  
  Now suppose we are in case \ref{case 2}.  By the construction of the $\mathbb{L}$-shape for column $\ell$, in $\hat{D}$ there must also be a $0$ in box $(j, \ell)$ (as it could not have been a last $\leplus$ in its row as there is still a $\leplus$ at $(j,o)$ with $o > \ell$) and the first $\leplus$ in column $\ell$ must be in some row $f < j$ (as otherwise the choice of $b_T$ would mean there is no $\leplus$ in box $(i, \ell)$ in $D$). 
  But now we are done as there is a \Le-configuration in $\hat{D}$ at $(j, o)$, $(j, \ell)$ and $(f, \ell)$, which is impossible as $\hat{D}$ is a \Le-diagram.
\end{proof}

\section{Proof of Le diagram T-duality}\label{S:proof}

Finally, we show that the construction described in Section~\ref{S:visual} gives the correct \Le-diagram and hence is the T-duality map.
We have that $D$ is a \Le-diagram of type $(k+1, n)$ from Theorem~\ref{Dlediag}. It remains to show that $D$ is loopless, has the right dimension, and has $\pi$ as its associated decorated permutation.

\bpoint{Dimension and looplessness of \texorpdfstring{$D$}{D}}\label{SS:dim}

A benefit of our approach is that we can clearly see how the dimension of $D$ arises in relation to the dimension of $\hat{D}$.
Namely, it comes from having at least one $\leplus$ in every column of $D$, with the number of additional $\leplus$'s being exactly the number of non-last $\leplus$'s in $\hat{D}$.

\tpointn{Theorem}\label{dimension}
\statement[eq]{
  Under the filling described in Section~\ref{S:visual}, $D$ is loopless and has dimension 
  \[\dim(S_D) = \dim(S_{\hat{D}}) - 2k + (n-1).\]
}

\begin{proof}
For this proof, we only need Proposition~\ref{number} and refer to (i) and (ii) from that statement.

First, it follows automatically from having $\geq 1$ $\leplus$ in every column of $D$ that $D$ is loopless.
For the dimension of $S_D$, \ref{plusi} and the $+1$ in \ref{plusii} gives one $\leplus$ in every column of $D$ for a total of $n - (k+1)$ $\leplus$'s.
Summing up what's left over in \ref{plusii}, using the $s_\ell$ and $t_\ell$ notation from Proposition~\ref{number}, gives:
\begin{align*}
\sum_{\ell} s_{\ell} + t_{\ell}
  &= \left( \# \text{ of non-last $\leplus$'s in $\hat{D}$ } - \# \text{ of non-last $\leplus$'s in col. $a_n$ of $\hat{D}$ }\right) \\[-1em]
  &\quad\quad + \left( \# \text{ of non-last $\leplus$'s in col. $a_n$ of $\hat{D}$ }\right)  \\
  &= \# \text{ of non-last $\leplus$'s in $\hat{D}$ } \\
  &= \dim(S_{\hat{D}}) - k
\end{align*}
where the sums run over columns $\ell \neq a_n$ in $\hat{D}$ with at least one $\leplus$.
The equalities come from:
\begin{itemize}
\item Summing over $s_{\ell}$ gives the number of non-last $\leplus$'s in each column in $\hat{D}$ excluding column $a_n$.
\item Summing over $t_{\ell}$ gives the number of rows with a $\leplus$ in column $a_n$ in $\hat{D}$ with last $\leplus$'s not in column $a_n$, which in other words is the number of non-last $\leplus$'s in column $a_n$.
\item As $\hat{D}$ is co-loopless and there are $k$ rows all with at least one $\leplus$, there are $k$ last $\leplus$'s out of a total of $\dim(S_{\hat{D}})$, which gives us the last equality.
\end{itemize}

Putting the two together gives $\dim(S_D) = \dim(S_{\hat{D}}) - 2k + (n-1)$ as needed:
\begin{align*}
\# \text{ of $\leplus$'s in $D$}
  &= \sum_{\text{cols in (i)}} 1 + \sum_{\text{cols in (ii)}} s_{\ell} + t_{\ell} + 1 
  = n - (k+1) + \dim(S_{\hat{D}}) - k 
  = \dim(S_{\hat{D}}) - 2k + (n-1)
  \tag*{\qedhere}
\end{align*}
\vspace{-3pt}
\end{proof}

\bpoint{\texorpdfstring{$D$}{D} corresponds to the correct decorated permutation}

Recall that $\hat{D}$ is a co-loopless \Le-diagram of type $(k,n)$ associated to the co-loopless decorated permutation $\hat\pi$ with $k$ anti-excedances, where
\[\hat\pi = \left(\begin{array}{cccc} 1 & 2 & \cdots & n \\ a_n & a_1 & \cdots & a_{n-1}\end{array}\right). \]

\tpointn{Theorem}\label{permutation}
\statement{
  Under the filling described in Section~\ref{S:visual}, the associated decorated permutation to $D$ is
  \[\pi = \left(\begin{array}{cccc} 1 & 2 & \cdots & n \\ a_1 & a_2 & \cdots & a_{n}\end{array}\right), \]
  where $\pi$ is loopless and has $k+1$ anti-excedances.
}

\begin{proof}
Since $D$ is loopless by Theorem~\ref{dimension} and is a \Le-diagram of type $(k+1,n)$ by Theorem~\ref{Dlediag}, its associated decorated permutation $\pi$ is also loopless and will have $k+1$ anti-excedances.
Now, what we want to prove is that
  \[ \begin{cases} \pi(i) = \hat\pi(i+1) = \begin{cases} a_i \neq i+1 & \text{ for non-fixed points $i+1$ of $\hat\pi$ } \\ i+1 & \text{ for fixed points $i+1$ of $\hat\pi$ } \end{cases} & \text{ for } i \neq n \\ \pi(n) = \hat\pi(1) = a_n & \end{cases} \]

To get from $D$ to its decorated permutation, we go through its pipe dream (see Section~\ref{SS:le-def}), that is starting from the label $i$ on the SE border, we follow its pipe until it reaches a label $j$ on the NW border of $D$, indicating that $\pi(i) = j$.
We will refer to the pipe starting at $i$ as the corresponding path for $\pi(i)$, or just for $i$, where the turns are indicating where the $\leplus$'s are.

In the following figures, the shaded areas indicate boxes filled with 0's. 

First, for the easiest case of $\pi(n)$, we look at the possibilities of what $n$ corresponds to in $D$:
\begin{figure}[h]
  \centering
  \begin{subfigure}[t]{0.25\textwidth}
      \centering\permproofn
      \caption{$n$ is a row in $D$}
  \end{subfigure}
  \begin{subfigure}[t]{0.3\textwidth}
      \centering \permproofncolzero
      \caption{$n$ is a column in $D$\\ and a column of all $0$'s in $\hat{D}$}
  \end{subfigure}
  \begin{subfigure}[t]{0.35\textwidth}
      \centering\permproofncol
      \caption{$n$ is a column in $D$\\ and a column with at least one $\leplus$ in $\hat{D}$}
  \end{subfigure}
\end{figure}

In case (a), as $\hat{D}$ is co-loopless, $n$ must be a column in $\hat{D}$ and since the only change in rows and columns is through $a_n$, we have $\pi(n) = n = a_n$ as needed.
In case (b), Theorem~\ref{glueing} tells us there is a horizontal string of $\leplus$'s at the end of diagram $D$ either glued to a row, in which case this row must be $a_n$ as $\hat{D}$ is co-loopless, or to the vertical part of the last $\mathbb{L}$ in $D$. 
Now each $\leplus$ in this vertical part of the $\mathbb{L}$ comes either from a last $\leplus$ in $\hat{D}$ in a row with a $\leplus$ in column $a_n$, or it's an extra $\leplus$ in row $a_n$.
However for a row $b$ to have a $\leplus$ at $(b, a_n)$ in $\hat{D}$, we must have $a_n > b$.
That is, this $\mathbb{L}$ has its bottom-most $\leplus$ in the vertical part in row $a_n$.
In either case, the $\leplus$ in column $n$ in $D$ is in row $a_n$ giving $\pi(n) = a_n$ as needed. 
Finally in case (c), Definition~\ref{shapes} places in $D$ an $\mathbb{L}$-shape whose vertical part is in column $n$.
Once again as this is the last $\mathbb{L}$ in $D$, with the same argument as in case (b), its bottom-most $\leplus$ in column $n$ is in row $a_n$ and thus $\pi(n) = a_n$.

Similarly, for the case where $i+1$ is a fixed point of $\hat\pi$ for $i \neq n$, so $i+1 \neq a_n$ is a column of all 0's in $\hat{D}$, we look at what $i$ corresponds to in $D$:
\begin{figure}[h]
  \centering
  \begin{subfigure}[t]{0.25\textwidth}
      \centering\permprooffixedrow
      \caption{$i$ is a row in $D$}
  \end{subfigure}
  \begin{subfigure}[t]{0.3\textwidth}
      \centering \permprooffixedcolzero
      \caption{$i$ is a column in $D$\\ and a column of all $0$'s in $\hat{D}$}
  \end{subfigure}
  \begin{subfigure}[t]{0.35\textwidth}
      \centering\permprooffixedcol
      \caption{$i$ is a column in $D$\\ and a column with at least one $\leplus$ in $\hat{D}$}
  \end{subfigure}
\end{figure}

For cases (a) and (b), regardless if the $\leplus$'s are part of an $\mathbb{L}$ or a string of $\leplus$'s, By Definition/Theorem~\ref{glueing} we have a $\leplus$ in box $(i, i+1)$ in case (a) or a string of $\leplus$'s in columns $i, i+1$ and in the same row in case (b). 
Either way we get $\pi(i) = i+1$ as needed.
In case (c), Definition~\ref{shapes} places an $\mathbb{L}$-shape with vertical part in column $i$ in $D$. 
By Theorem~\ref{glueing}, whichever shape is containing the $\leplus$ in column $i+1$ must glue to the left of this $\mathbb{L}$ at the last $\leplus$ in column $i$, say at $(b_L, i)$.
Thus we get $\leplus$'s at $(b_L, i)$ and $(b_L, i+1)$ with all 0's below in column $i$ and above in column $i+1$, which gives $\pi(i) = i+1$ as needed.

Lastly, we have the hardest case of when $i+1$ is not a fixed point of $\hat\pi$ for $i \neq n$, that is in $\hat{D}$ either $i+1$ is a column with at least one $\leplus$ or $i+1$ is a row.
Say $\hat{\pi}(i+1) = j$ for some $j \neq i+1$ (we also have $j \neq a_n$).
We want to show that $\pi(i) = j$.
In the subsequent proofs, Definitions~\ref{shapes} and \ref{strings} and Definition/Theorem~\ref{glueing} will be used without explicit citation.

Note that in general, every path in $\hat{D}$ for a non-fixed point $i+1$ must start with one $\leplus$ in row/column $i+1$ and then is built from alternating between two $\leplus$'s in the same row and two in the same column, or vice versa, until row/column $j$ where the path ends with one $\leplus$ in row/column $j$.
Consider the following path in $\hat{D}$ which goes through $m+1$ columns of $\leplus$'s,
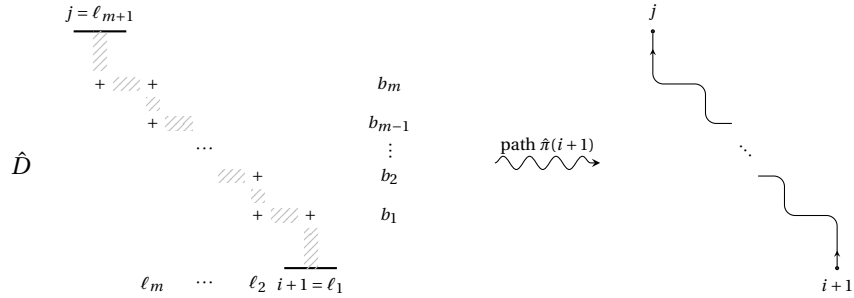
\[ \permproofgeneralpath \]
where $\ell_j$ denotes the columns of $\leplus$'s and $b_j$ denotes the rows of $\leplus$'s.
Here, note that $i+1 > b_1 > \cdots > b_m$ and $\ell_1 < \cdots < \ell_{m+1}$.
If instead $i+1$ is a row, there is an additional $\leplus$ in the first column of $\leplus$'s at $(i+1, \ell_1)$ where $i+1 > b_1$.
If instead $j$ is a row, there is an additional $\leplus$ in the last column of $\leplus$'s at $(j, \ell_{m+1})$ where $j < b_m$.
Since the shaded areas in the path are filled with all $0$'s, by the \Le-condition for $\hat{D}$, the vertical shaded regions extend to the left until the end of the diagram and the horizontal shaded regions extend to the top of the diagram.

Now to tackle the problem at hand, we split the paths $\hat\pi(i+1)$ in $\hat{D}$ based on its relation to $a_n$ and consider them separately.
As an example of the different types of paths, see Figure~\ref{fig:permpathex}.
\begin{enumerate}
\item $i+1 < j < a_n$: the path in $\hat{D}$ is completely to the right of $a_n$ and $j$ is a column.
\item $j < i+1 \leq a_n$ or $i+1 \leq a_n < j$: the path passes through column $a_n$ and thus either the last $\leplus$ in the path is before $a_n$ (and $j$ is a row), or the path contains $\leplus$'s in column $a_n$.
\item $j < a_n < i+1$ or $a_n < i+1 < j$: the path passes through where row $a_n$ in $D$ will be.
\item $a_n < j < i+1$: the path is completely below where row $a_n$ in $D$ will be and $j$ is a row.
\end{enumerate}

\begin{figure}[h]
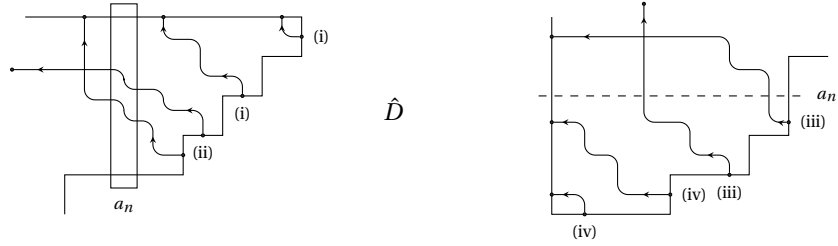

\centering
\[ \permproofpathex \]
\caption[Examples of the four types of paths in $\hat{D}$ considered in the proof of $\pi$.]{Examples of the four types of paths in $\hat{D}$ considered in the proof of $\pi$. Each path is labelled by its type. On the left, the rectangle indicates where the column $a_n$ is in $\hat{D}$. On the right, the dashed line indicates where the row $a_n$ will be in $D$.}
\label{fig:permpathex}
\end{figure}

Note all the paths in (ii) with $\leplus$'s to the left of $a_n$ must contain $\leplus$'s in column $a_n$ because of the \Le-condition for $\hat{D}$, since we know there is a $\leplus$ at $(1, a_n)$ whenever a box exists there.
In the special case of $a_n = 1$, paths in (i) and (ii) do not exist.

Starting with (i) and (iv), we are in the cases where in $\hat{D}$ we have, respectively:
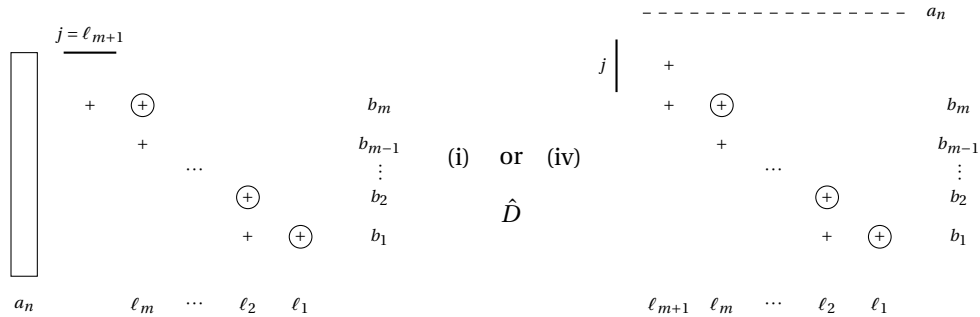
\[ \permproofihat \]
and either $i+1 = \ell_1$ is a column or $i+1 > b_1$ is row with a $\leplus$ at $(i+1, \ell_1)$.
Notice that the top $\leplus$ in each column in the path, except possibly $\ell_{m+1}$, is not the last $\leplus$ in their respective rows in $\hat{D}$.
In the figure above, these are the circled $\leplus$'s.
This implies that in the $\mathbb{L}$'s corresponding to these columns in $D$, there is also a $\leplus$ in these positions, $(b_1, \ell_1), \ldots (b_m, \ell_m)$; these $\leplus$'s are circled in the figures below.

For cases (i) and (iv), since $a_n > j$ or $a_n < j$, all $\mathbb{L}$-shapes corresponding to columns $\ell_1, \ldots, \ell_m$, and for (i) also column $j$, each have their $b_T = h \neq f, a_n$ where $f$ is the row of the first $\leplus$ in $\hat{D}$ in that column. 
Now let's look at two consecutive columns $\ell_s, \ell_{s+1}$, for $s \neq m$.
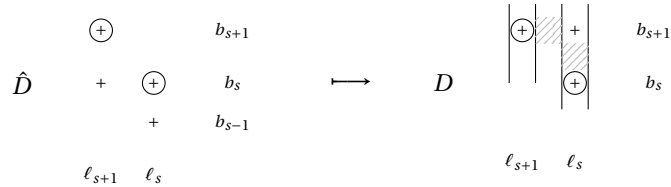
\[ \permprooficols \]

We know that any $\leplus$'s in column $\ell_s$ strictly between rows $b_s, b_{s+1}$ must be last $\leplus$'s in their row because of the \Le-condition for $\hat{D}$ and thus these correspond to a $0$ at $(b, \ell_s)$ in $D$.
If there are $\leplus$'s in column $\ell_s$ above $b_{s+1}$ in $\hat{D}$, then by the \Le-condition there must be a $\leplus$ at $(b_{s+1},\ell_s)$ in $\hat{D}$.
In particular, this is not the last $\leplus$ in row $b_{s+1}$ and thus there is a $\leplus$ at $(b_{s+1}, \ell_s)$ in $D$.
If the first $\leplus$ in column $\ell_s$  in $\hat{D}$ is in row $b_{s+1}$ or below, since there are no $\leplus$'s in rows $b$ for $b_s > b > b_{s+1}$ after column $\ell_s$ by the \Le-condition for $\hat{D}$ applied to row $b_s$, then row $b_{s+1}$ is the first row above $b_s$ that has a $\leplus$ to the left of column $\ell_s$.
That is, we have that $b_T = b_{s+1}$ (by the $h$-condition in Definition~\ref{shapes}) for the $\mathbb{L}$ in column $\ell_s$ and thus there is a $\leplus$ at $(b_{s+1}, \ell_s)$.
In either case, the $\mathbb{L}$ in column $\ell_s$ in $D$ always has a $\leplus$ at $(b_{s+1}, \ell_s)$ and at $(b_s, \ell_s)$, with 0's between.
Since in $D$ the $\mathbb{L}$ in column $\ell_{s+1}$ glues to the left of the $\mathbb{L}$ in column $\ell_{s}$ at its bottom-most $\leplus$, there must be $0$'s in row $b_{s+1}$ strictly between columns $\ell_s$ and $\ell_{s+1}$.
Therefore, we get the above figure on the right.

For the final columns $\ell_{m}, \ell_{m+1}$ we look at the two path cases separately. 
For (i) and column $j=\ell_{m+1}$, notice that in $\hat{D}$, there must be a row $b < b_m$ with a $\leplus$ to the left of column $j$ since we know at the least there's a $\leplus$ at $(1, a_n)$.
Since we know that $(b_m,j)$ is the first $\leplus$ in column $j$ and that $b_T \neq j, a_n$ for the $\mathbb{L}$ in column $j$ in $D$, we must have that $b_T = b$ and thus there is a $\leplus$ at $(b, j)$ with 0's all above in $D$.
For column $\ell_m$, using the same argument as for $\ell_s$ with row $b$ as $b_{s+1}$, in the $\mathbb{L}$ in column $\ell_m$, there must be a $\leplus$ at $(b, \ell_m)$ and at $(b_m, \ell_m)$ with 0's in-between.

For (iv) and columns $ \ell_m, \ell_{m+1}$, we have the same analysis as for $\ell_s$, $\ell_{s+1}$ except that now the $\leplus$ at $(j,\ell_{m+1}$) is the last $\leplus$ of its row in $\hat{D}$, which is a row without a $\leplus$ in column $a_n$, and thus corresponds to a $0$ at $(j, \ell_{m+1})$ in the $\mathbb{L}$-shape for column $\ell_{m+1}$.

Now, since we know that $\mathbb{L}$'s glue to the left of previous $\mathbb{L}$'s at the first $\leplus$ from the bottom, we get the following in $D$, regardless if $i+1$ is a column or a row:
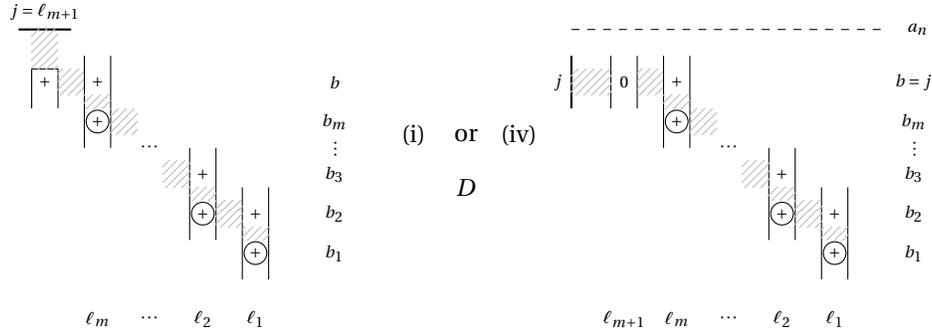
\[ \permproofi \]
Therefore, the path that starts at $(b_1, \ell_1)$ ends at column $\ell_{m+1}$ for paths in (i) or row $b$ for (iv) respectively and hence in either case ends at $j$.

Finally to look at what happens with $i$ in $D$.
\begin{figure}[h]
  \centering
  \begin{subfigure}[b]{0.3\textwidth}
      \centering\permproofibothcols
      \caption{$i, i+1$ are columns in $D$}
  \end{subfigure}
  \begin{subfigure}[b]{0.3\textwidth}
      \[\permprooficolrow\]
      \[\centering\permproofibothrows\]
      \caption{$i$ is a row in $D$}
  \end{subfigure}
  \begin{subfigure}[b]{0.35\textwidth}
      \centering\permproofirowcol
      \caption{$i$ is a column, $i+1$ is a row in $D$}
  \end{subfigure}
\end{figure}

In case (a), regardless of how many $\leplus$'s are in column $i$ in $\hat{D}$, we must have in $D$ a $\leplus$ at $(b_1, i)$ and 0's below in column $i$ since we know how $\mathbb{L}$'s glue.
In case (b), as $\hat{D}$ is co-loopless and by the \Le-condition for $\hat{D}$, we must have that $b_1 = i$ (when $i+1$ is also a row, recall that there are $\leplus$'s at $(i+1,\ell_1), (b_1,\ell_1)$ in $\hat{D}$).
In case (c), notice $b_1 < i$ and is a row in $\hat{D}$ with its last $\leplus$ to the left of column $i$.
Then in the first column on or to the right of $i$ with at least one $\leplus$ in $\hat{D}$, any $\leplus$'s below row $b_1$ must be last $\leplus$'s in their row by the \Le-condition for $\hat{D}$. 
Thus the corresponding $\mathbb{L}$ to this column would have its bottom-most $\leplus$ in row $b_1$ by the \Le-condition for $D$. 
If there are no such columns, as $\hat{D}$ is co-loopless, row $b_1$ must be the bottom-most row of column $i$. 
In particular, by these properties and as we know how shapes glue, whichever shape is in column $i$ in $D$ must have a $\leplus$ at $(b_1, i)$ with all $0$'s below in column $i$.

In all the cases, we can connect these path starts at the circled point to the rest of the path, and thus we get that the corresponding path to $i$ in $D$ leads to $j$, giving us $\pi(i) = j$ as needed for paths in (i) and (iv).

Now consider cases (ii) and (iii).  For paths in (ii), first notice we already have the case when the last $\leplus$ in the path is in a column before $a_n$ and $j$ is a row. 
In this case, all the arguments from paths in (i) stay the same except for the path end for which we can use the argument from paths in (iv).

When the path in (ii) contains $\leplus$'s in column $a_n$ or in (iii) it passes through where row $a_n$ will be, we have in $\hat{D}$ respectively:
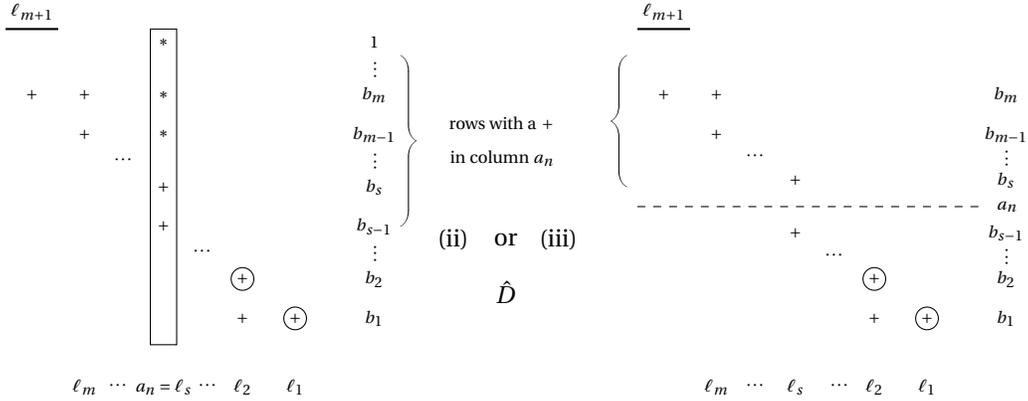
\[ \permproofiihat \]
where for (ii) $\ell_s=a_n$ and for (iii) $\ell_s$ is the column whose $\leplus$'s straddle the row that will be $a_n$, and where $*$ is indicating $\leplus$'s in column $a_n$ that are not part of the path and the dashed line refers to where row $a_n$ will be in $D$. 
We also have either $j = \ell_{m+1}$ is a column or $j < b_m$ is row with a $\leplus$ at $(j, \ell_{m+1})$, and either $i+1 = \ell_1$ is a column or $i+1 > b_1$ is row with a $\leplus$ at $(i+1, \ell_1)$.

Then notice for columns $\ell_1, \ldots, \ell_{s-1}$, the analysis from paths in (i) and (iv) stay exactly the same.
For columns $\ell_{s+1}, \ldots, \ell_{m+1}$, we note that all rows $b_{s}, \ldots, b_m$, and $j$ if it is a row, must have $\leplus$'s in column $a_n$ in $\hat{D}$ by the \Le-condition as there is a $\leplus$ at $(1, a_n)$.
In particular, this means that the $\mathbb{L}$'s corresponding to columns $\ell_{s+1}, \ldots, \ell_{m+1}$, which are all $> a_n$, have $\leplus$'s exactly in the same positions as in $\hat{D}$ in rows $< a_n$.
For paths in (ii), we are done with all the relevant columns as to get $D$, column $a_n=\ell_s$ is removed and we know how the $\mathbb{L}$'s are glued.

For paths in (iii), we need to look at what happens to column $\ell_s$, which row $a_n$ will pass through, and column $\ell_{s-1}$.
For column $\ell_s$, since row $a_n$ will occur strictly between rows $b_s$ and $b_{s-1}$, there will be a $\leplus$ at $(a_n, \ell_s)$ in the $\mathbb{L}$ corresponding to column $\ell_s$.
Since row $b_s$ has a $\leplus$ in column $a_n$ in $\hat{D}$, with $\ell_s > a_n$, there is also a $\leplus$ at $(b_s, \ell_s)$ in the $\mathbb{L}$.

For column $\ell_{s-1}$, we already have a $\leplus$ at $(b_{s-1}, \ell_{s-1})$ in its corresponding $\mathbb{L}$. 
Now, if the first $\leplus$ in column $\ell_{s-1}$ in $\hat{D}$ is in a row $< a_n$ then the corresponding $\mathbb{L}$ will contain row $a_n$ and thus there will be a $\leplus$ at $(a_n, \ell_{s-1})$. 
Otherwise, if the first $\leplus$ is at $(b, \ell_{s-1})$ for $b_{s-1} \geq b > a_n$ in $\hat{D}$ then the corresponding $\mathbb{L}$ will have $b_T = a_n$, and a $\leplus$ at $(a_n, \ell_{s-1})$, as by the \Le-condition for $\hat{D}$ there cannot be any rows in-between $b_{s-1}$ and $a_n$ with a $\leplus$ to the left of column $\ell_{s-1}$.
Additionally, this means any $\leplus$'s at $(b, \ell_{s-1})$ in $\hat{D}$, for $b_{s-1} > b > a_n$ must be last $\leplus$'s in its row, which does not have a $\leplus$ in column $a_n$ in $\hat{D}$, and corresponds to $0$'s in the $\mathbb{L}$ for column $\ell_{s-1}$.
That is, the $\mathbb{L}$ in column $\ell_{s-1}$ always has $\leplus$'s at $(a_n, \ell_{s-1})$ and $(b_{s-1}, \ell_{s-1})$ with 0's between. 

Thus we get the following in $D$ regardless if $i+1$ or $j$ is a column or a row:
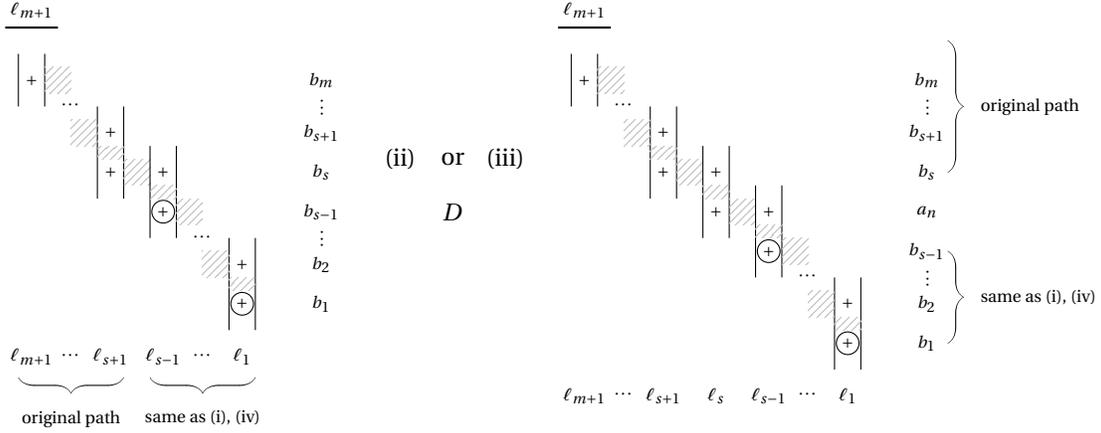
\[ \permproofii \]

For the path ends, as columns after and rows above $a_n$ are following the original paths from $\hat{D}$, when $j = \ell_{m+1}$ is a column, and $a_n > b_m$, there are only $0$'s above row $b_m$ in column $\ell_{m+1}$ in $D$ and thus the path $\pi(i)$ turns up after $(b_m, \ell_{m+1})$.
When $j = b < b_{m}$ is a row, there is an extra $\leplus$ at $(j, \ell_{m+1})$ and all $0$'s to the left in row $j$ in $D$, and thus the path $\pi(i)$ turns up at $(b_m, \ell_{m+1})$ and then to the left at $(j, \ell_{m+1})$.

In the special case of $a_n < b_m$ where $j = \ell_{m+1}$, since we know that the $\leplus$ at $(b_m, j)$ is the first in its column in $\hat{D}$, the $\mathbb{L}$ in column $j$ in $D$ must have its $b_T = a_n$ with a $\leplus$ at $(a_n, j)$.
Column $\ell_{m}$ stays the same as before with $\leplus$'s at $(a_n, \ell_m)$ and $(b_m, \ell_m)$, and 0's between, in $D$ and so we get:
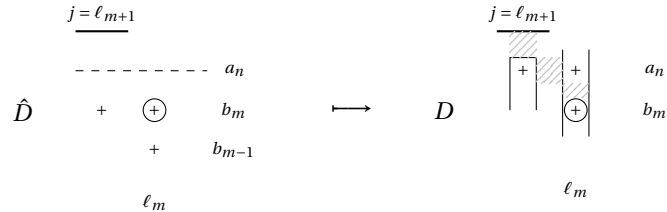
\[ \permproofiiispecialone \]

In all of the cases, the portion of the path $\pi(i)$ starting at $(b_1, \ell_1)$ will end at $j$ in $D$ as needed.

Finally, what's left is to look at the part starts. 
The analysis on $i+1$ and $i$ stays exactly the same as before (since it only depends on $\hat{D}$), with the slight exception for paths in (ii) when $i+1 = \ell_1 = a_n$, and for paths in (iii) when $a_n > b_1$.
When $i+1 = \ell_1 = a_n$, the only difference is that the circled $\leplus$ in the analysis for $i$ in $D$ is now the $\leplus$ at $(b_1, \ell_2)$, and we consider $i+1$ to be a row.
Thus we get that $\pi(i) = j$ as needed for paths in (ii). 

In the special case of $a_n > b_1$ where $i+1 = \ell_1$, now the $\mathbb{L}$ in column $\ell_1$ in $D$ will have a $\leplus$ at $(a_n, \ell_1)$, since $\ell_1 > a_n$ and thus its $b_B = a_n$.
There is also still a $\leplus$ at $(b_1, \ell_1)$, as there's a non-last $\leplus$ there in $\hat{D}$.
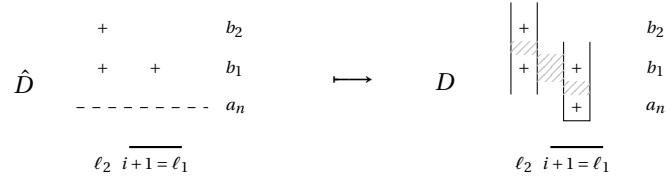
\[ \permproofiiispecialtwo \]
Now $a_n$ plays the role of $b_1$ and we use the same arguments as before, even if $i = a_n$ in which case we consider $i$ to be a row.
In all cases, we get that $\pi(i) = j$ as needed for paths in (iii).

Putting (i)--(iv) together, we are now done since for all paths $\hat\pi(i+1) = j$ we get $\pi(i) = \hat\pi(i+1) = j$.
\end{proof}

\bpoint{T-duality on the level of \texorpdfstring{\BLe-}{Le }diagrams}

With the proof that $D$ gives the correct decorated permutation complete, we now put everything together.
Theorems~\ref{Dlediag},~\ref{dimension} and~\ref{permutation}, tell us that the construction given in Section~\ref{S:visual} is the T-duality map (see Section~\ref{S:t-duality}) on the level of \Le-diagrams.

\tpointn{Theorem}\label{letheorem}
\statement{
  Given a co-loopless \Le-diagram $\hat{D}$ of type $(k,n)$, the map defined in Section~\ref{S:visual} constructs a loopless \Le-diagram $D$ of type $(k+1,n)$ such that
  \begin{itemize}
    \item the dimensions of the positroid cells indexed by $D$ and $\hat{D}$ relate via
      \[ \dim(S_D) = \dim(S_{\hat{D}}) - 2k + (n-1), \]
    \item and the associated decorated permutations to $D$ and $\hat{D}$ relate via T-duality.
  \end{itemize}
  That is, we have given the T-duality map from $\hat{\pi} \mapsto \pi$ on the level of \Le-diagrams.
}

\section{Discussion}\label{S:discussion}

We gave an explicit construction for the T-duality map at the level of $\Le$-diagrams.  It is a particularly convenient construction for understanding the change in dimension under T-duality.  Specifically, we see that we get at least one $\leplus$ in every column of $D$, with the number of additional $\leplus$'s being exactly the number of non-last $\leplus$'s in $\hat{D}$, which is exactly the expected relationship, see Section~\ref{SS:dim}.
There are a couple of extensions and other perspectives that one can consider, which we outline here.

We can also see the inverse map from this perspective.  The shape of $\hat{D}$ gives the shape of $D$ by reversing Figure~\ref{fig:construct-D}. For the filling, by the second point of Remark~\ref{rem L rems}, the vertical part of each $\mathbb{L}$ shape has height at least 2 and so any \Le-diagram $D$ with at least one $\leplus$ in every column can be decomposed uniquely into $\mathbb{L}$ shapes with potentially also one string of $\leplus$'s on the left side of any section.
 Then from the positions of the $\leplus$'s in the vertical part of each $\mathbb{L}$ the positions of the $\leplus$'s can be read off, reversing the original construction of the $\mathbb{L}$'s (Definition~\ref{shapes}).

We can also consider iterating the T-duality map.
If $D$ is co-loopless, that is $\pi$ has no fixed points, then the construction in Section~\ref{S:visual} can be applied again, this time starting with $D$, to obtain a \Le-diagram $\bar{D}$ whose associated decorated permutation would be
\[\bar\pi = \left(\begin{array}{cccc} 1 & \cdots & n-1 & n \\ a_2 & \cdots & a_{n} & a_{1} \end{array}\right). \]
That is from the original $\hat\pi$, the permutation is shifted to the left twice.
We know that the shape of $\bar{D}$ corresponds to removing the column $a_1 = \pi(1)$ from $D$ and adding in a row labelled $a_1$ in the appropriate place.
For the filling of $\bar{D}$, from Theorem~\ref{glueing} we know that we really only need to look at what happens to the $\mathbb{L}$-shapes.
Applying Definition~\ref{shapes} to an $\mathbb{L}$-shape gives
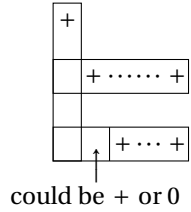
\[ \doublel \]
where now every $\leplus$ in the horizontal string of $\leplus$'s turns into its own $\mathbb{L}$, just with no horizontal part. 

As a purely combinatorial map on decorated permutations, T-duality (Definition~\ref{def:t-duality}) also appears as a special case of a more general operation studied in~\cite{BCT} (the $A = \emptyset$ case of Definition 23; note their decorations for fixed points are used opposite of ours).
For this map, looking at the left-shift $\hat{\pi} \to \pi$ direction, now $\hat{\pi}$ no longer needs to be co-loopless and we can specify a set $A$ of positions $i$ to "freeze", i.e. for $i \in A$, $\pi(i) = \hat{\pi}(i)$.

For our \Le-diagram construction, co-loops $i$ in $\hat{\pi}$ can be dealt with as if they were originally loops. 
The only change is in the shape of $D$ as now $i$ is a column where previously $i$ was a row in $\hat{D}$. 
For freezing fixed points $i \neq 1,n$, which are then decorated as co-loops in $\pi$, the operation turns into simply adding to $D$ a row $i$, in the correct position, of all $0$'s (and if it was previously a column, removing column $i$).
The more interesting operation would be freezing non-fixed points or when positions $1$ or $n$ are frozen. 

Finally, going back to the motivation for studying the T-duality map, one of the key ingredients in~\cite{PSBW} for proving the correspondence between tilings of the hypersimplex and of the amplituhedron came down to looking at T-duality as a map on plabic graphs. 
Formulated in the original $\pi \to \hat{\pi}$ direction, this map turned out to be a particularly nice construction on graphs, similar to that of taking a dual, see Definition 8.7. 
From our \Le-diagram construction, and via its bijection to plabic graphs (see \textsection 20 of~\cite{tpgrass}), we can already see how the horizontal strings of $\leplus$'s correspond to black lollipops (loops in $\hat{\pi}$). 
As each $\leplus$ in the string maps to a trivalent white vertex, the entire horizontal string corresponds to a sequence of connected white vertices, each also connected to the boundary of the plabic graph.
Thus these white vertices are enclosing boundary faces. 
By Definition 8.7 in~\cite{PSBW}, a black vertex is placed in each of these faces and only connected to the boundary giving the black lollipops as needed. 
What would be interesting is if we can directly interpret the rest of the T-duality on plabic graphs map via our $\mathbb{L}$-shapes.

\appendix
\section{Algorithmic row-by-row reformulation}\label{S:algorithm}

We can reformulate our characterization of the T-duality map at the level of \Le-diagrams by writing an explicit algorithm that acts row by row.

Using definitions and notation from before, the shape of $D$ is still as described in Section~\ref{SS:shape}.  It remains then to fill $D$ with $0$'s and $\leplus$'s.  
We will fill $D$ row by row, from right to left, based on the corresponding row and rows below in $\hat{D}$.
For row $a_n$, since $a_n$ was a column in $\hat{D}$, we will consider its "corresponding row" in $\hat{D}$ as a row of all 0's of the same length as in $D$ and placed in the same position as $D$ (in-between rows $b_{j-1}$ and $b_j$ so that the order of labels of the SE border is maintained).

Let $L_u$ be the column containing the leftmost $\leplus$ in each row labelled $b_u$ of $\hat{D}$ (which is in box $(b_u, L_u)$) for $1 \leq u \leq k$.
We'll take the convention that $L_n > n$ as with this convention the algorithm does not need a special case for row $a_n$.
Let $W_u$ be the row directly below row $b_u$, and let $W_n$ be the row directly below $a_n$, if they exist.
Otherwise, for the last row $b_k$, define $W_k=n+1$, or if $a_n$ is the last row, then define $W_n=n+1$.
Concretely we have
\[ W_u = \begin{cases} b_{u+1} & 1 \leq u < k, u \neq j-1 \\ a_n & u = j-1 \\ b_{j+1} & u = n, j \neq k+1 \\ n+1 & u=k, j \neq k+1 \text{ or } u=n, j=k+1 \end{cases} \]
In particular, $W_u - 1$ is the column right before the next row below, or for the last row, $W_u-1=n$.
Finally, we say that a $0$ is \textbf{restricted} if there is a $\leplus$ in a box to its left, in the same row, and \textbf{unrestricted} otherwise.  By convention we take the leftmost $\leplus$ of row $a_n$ to be at $(a_n, L_n)$ with $L_n>n$.
In Example~\ref{ex:notation}, the only restricted $0$ is in box $(1,2)$.

There are two different row types to consider for $D$:
\begin{enumerate}[label=(\Roman*)]
  \item Rows $b_u$ such that the box $(b_u, a_n)$ either does not exist, or does not have a $\leplus$ in $\hat{D}$.
  \item Rows $b_u$ such that the box $(b_u, a_n)$ has a $\leplus$ in $\hat{D}$.
    Note: These rows will always be above row $a_n$.
\end{enumerate}

\tpointn[n]{Algorithm for filling in rows of $D$}\label{le-alg-main}{ }
Run in parallel for each row:

\fbox{
\hspace{2.3em}
\begin{minipage}[t]{.92\linewidth}
\hspace{-2.3em}\textbf{For row $b_u$ of type (I):}
Fill boxes in row $b_u$ from right to left, if they exist, as follows:
\begin{enumerate}[label=Step \arabic*.,itemsep=4pt]
  \item For columns $b_u+1$ to $W_u-1$, fill boxes in row $b_u$ with $\leplus$'s,\\
    i.e. fill boxes $(b_u, b_u+1), \ldots, (b_u, W_u-1)$ with $\leplus$'s.
  \item For columns $W_u+1$ to $L_u-1$, fill boxes in row $b_u$ with $\leplus$'s under the conditions defined below.
    Otherwise, fill the boxes with 0's.\\
    i.e. fill boxes $(b_u, W_u+1), \ldots, (b_u, L_u-1)$ with $\leplus$'s if they satisfy the conditions below.
  \item Starting from column $L_u$, fill the rest of the boxes in row $b_u$ with $0$'s,\\
    i.e. fill boxes $(b_u, L_u)$ and leftwards to the end of the row with $0$'s. \\
\end{enumerate}
Note: if $W_u  > L_u$, then Step 1 only fills boxes until column $L_u-1$ and Step 2 is skipped.    
\vspace{3pt}
\end{minipage}
}

\fbox{
\hspace{2.3em}
\begin{minipage}[t]{.92\linewidth}
\hspace{-2.3em}\textbf{For row $b_u$ of type (II):}
Follow the algorithm for type (I) with the following modifications:
\begin{enumerate}[itemsep=4pt]
  \item[Steps 1 \& 2.] For columns $b_u+1$ until row $a_n$, follow Steps 1 \& 2 of row type (I), \\
    i.e. fill boxes $(b_u, b_u+1), \ldots, (b_u, a_n-1)$ as in row type (I).
  \item[Step 3.] Starting from column $a_n+1$, fill the rest of the boxes in row $b_u$ to agree with $\hat{D}$, \\
    i.e. fill boxes $(b_u, a_n+1)$ and leftwards until the end of the row according to the filling of box $(b_u, a_n+1)$ in $\hat{D}$.
\end{enumerate}  
\vspace{3pt}
\end{minipage}
}

\fbox{
\begin{minipage}[t]{.92\linewidth}
\textbf{Conditions for filling in Step 2:}
In Step 2, fill a box in row $b_u$ at $(b_u, \ell)$ with a $\leplus$ if:
\begin{enumerate}[itemsep=4pt]
  \item \label{cond 1} In $\hat{D}$, there is a $\leplus$ in box $(b_u, \ell)$.
  \item \label{cond 2} In $\hat{D}$, there is a $\leplus$ in some row below, say at $(b_m, \ell)$ where $m > u$, such that there are only unrestricted $0$'s in column $\ell$ in-between rows $b_u$ and $b_m$.\\
    Note: In the case that $m = u+1$, this condition holds trivially.
  \item \label{cond 3} In $\hat{D}$, column $\ell$ only has unrestricted $0$'s below row $b_u$.
      In other words in column $\ell$, all the rows below $b_u$ has their leftmost $\leplus$ before column $\ell$, i.e. their last $\leplus$ has already passed, where by convention the leftmost $\leplus$ of row $a_n$ occurs at $(a_n, L_n)$ for $L_n>n$.
      Note: This is a special case of \ref{cond 2}.
\end{enumerate}
\vspace{3pt}
\end{minipage}
\hspace{2.3em}
}

An example of the conditions is given in Figure~\ref{fig:step2-conds}.
Another way to phrase condition \ref{cond 2} is as follows: Look for $\leplus$'s in column $\ell$ in rows below $b_u$ where all rows in-between have no $\leplus$'s to the left of column $\ell$ (their leftmost $\leplus$ is before column $\ell$).

\begin{figure}[h]
  \centering
  \[ \steptwoii \]
  \caption[Example of conditions \ref{cond 2} and \ref{cond 3}.]{Example of conditions \ref{cond 2} and \ref{cond 3}. For row $b_u$, the columns of the two circled $\leplus$'s in this $\hat{D}$ satisfy condition \ref{cond 2} as all boxes above the $\leplus$ (until row $b_u$) are filled with unrestricted $0$'s. Thus in $D$, a $\leplus$ would be filled in row $b_u$ in those two columns.
  The column of the non-circled $\leplus$ not in row $b_u$ would fail this condition because of the row directly above it.
  The shaded columns indicate the columns that satisfy condition \ref{cond 3} since all the boxes below row $b_u$ are filled with unrestricted $0$'s.}
  \label{fig:step2-conds}
\end{figure}

See Section~5.1.5 of \cite{Hmmath} for a detailed worked example of applying the algorithm.

One interesting aspect of this algorithm is that for any particular row, only a part of $\hat{D}$ is looked at, namely it looks at the row itself and the first rows below for which the leftmost $\leplus$ has not yet passed.

The proof that this algorithm agrees with the approach via glueing $\mathbb{L}$'s can be sketched as follows.  First, we check that for each individual $\mathbb{L}$, the algorithm above builds the horizontal string of $\leplus$'s using Step 1 or Step 2 condition \ref{cond 3}.  Next, we check that the algorithm above builds the vertical part of each $\mathbb{L}$ using Step 3 and Step 2 conditions \ref{cond 1} and \ref{cond 2}.  Counting in each case, we can directly show that the number of $\leplus$'s in each vertical part of an $\mathbb{L}$ is $s_\ell + t_\ell + 1$.  The argument for the strings of $\leplus$'s is similar but simpler.  Finally, to show the algorithm above glues the $\mathbb{L}$'s in each section as we described in Theorem~\ref{glueing}, we can consider the possibilities for $b_B$.  For full details see Section 5.2 of \cite{Hmmath}.

\printbibliography

@article {LPW,
    author = {Lukowski, Tomasz and Parisi, Matteo and Williams, Lauren K},
    title = {The Positive Tropical Grassmannian, the Hypersimplex, and the m = 2 Amplituhedron},
    journal = {International Mathematics Research Notices},
    pages = {rnad010},
    year = {2023},
    month = {03},
    issn = {1073-7928},
    doi = {10.1093/imrn/rnad010},
    url = {https://doi.org/10.1093/imrn/rnad010},
    eprint = {2002.06164},
}

@article {PSBW,
    AUTHOR = {Parisi, Matteo and Sherman-Bennett, Melissa and Williams, Lauren K.},
     TITLE = {The {$m=2$} amplituhedron and the hypersimplex: {S}igns, clusters, tilings, {E}ulerian numbers},
   JOURNAL = {Comm. Amer. Math. Soc.},
  FJOURNAL = {Communications of the American Mathematical Society},
    VOLUME = {3},
      YEAR = {2023},
     PAGES = {329--399},
      ISSN = {2692-3688},
   MRCLASS = {05Exx (13F60 14M15 52B40)},
  MRNUMBER = {4612722},
       DOI = {10.1090/cams/23},
       URL = {https://doi.org/10.1090/cams/23},
    eprint = {2104.08254},
}

@article {tpgrass,
    AUTHOR = {Postnikov, Alexander},
     TITLE = {Total positivity, Grassmannians, and networks},
      YEAR = {2006},
       URL = {http://math.mit.edu/~apost/papers/tpgrass.pdf},
}

@inproceedings {post,
    AUTHOR = {Postnikov, Alexander},
     TITLE = {Positive {G}rassmannian and polyhedral subdivisions},
 BOOKTITLE = {Proceedings of the {I}nternational {C}ongress of
              {M}athematicians---{R}io de {J}aneiro 2018. {V}ol. {IV}.
              {I}nvited lectures},
     PAGES = {3181--3211},
 PUBLISHER = {World Sci. Publ., Hackensack, NJ},
      YEAR = {2018},
   MRCLASS = {05E18 (05B35 13F60 52B12 52B70)},
  MRNUMBER = {3966528},
    eprint = {1806.05307},
}

@book {grassampl,
    AUTHOR = {Arkani-Hamed, Nima and Bourjaily, Jacob and Cachazo, Freddy
              and Goncharov, Alexander and Postnikov, Alexander and Trnka,
              Jaroslav},
     TITLE = {Grassmannian geometry of scattering amplitudes},
 PUBLISHER = {Cambridge University Press, Cambridge},
      YEAR = {2016},
     PAGES = {ix+194},
      ISBN = {978-1-107-08658-6},
   MRCLASS = {81-02 (81R05 81R10 81T60)},
  MRNUMBER = {3467729},
MRREVIEWER = {Yasuhiro Abe},
       DOI = {10.1017/CBO9781316091548},
       URL = {https://doi.org/10.1017/CBO9781316091548},
    eprint = {1212.5605},
}

@article{ampl,
    author = {Arkani-Hamed, Nima and Trnka, Jaroslav},
     title = {The Amplituhedron},
   journal = {Journal of High Energy Physics},
    volume = {2014},
      year = {2014},
    number = {10},
 publisher = {Springer Science and Business Media LLC},
      ISSN = {1029-8479},
       DOI = {10.1007/jhep10(2014)030},
       url = {http://dx.doi.org/10.1007/JHEP10(2014)030},
    eprint = {1312.2007},
}

@article {GGMS,
    AUTHOR = {Gelfand, I. M. and Goresky, R. M. and MacPherson, R. D. and
              Serganova, V. V.},
     TITLE = {Combinatorial geometries, convex polyhedra, and {S}chubert
              cells},
   JOURNAL = {Adv. in Math.},
  FJOURNAL = {Advances in Mathematics},
    VOLUME = {63},
      YEAR = {1987},
    NUMBER = {3},
     PAGES = {301--316},
      ISSN = {0001-8708},
   MRCLASS = {14M15 (05B35 22E45 22E70 32C38 32C45 32M10)},
  MRNUMBER = {877789},
MRREVIEWER = {Hiroaki Terao},
       DOI = {10.1016/0001-8708(87)90059-4},
       URL = {https://doi.org/10.1016/0001-8708(87)90059-4},
}

@book {fulton,
    AUTHOR = {Fulton, William},
     TITLE = {Young tableaux},
    SERIES = {London Mathematical Society Student Texts},
    VOLUME = {35},
      NOTE = {With applications to representation theory and geometry},
 PUBLISHER = {Cambridge University Press, Cambridge},
      YEAR = {1997},
     PAGES = {x+260},
      ISBN = {0-521-56144-2; 0-521-56724-6},
   MRCLASS = {05E10 (05E05 05E15 14M15 20G05)},
  MRNUMBER = {1464693},
MRREVIEWER = {Tadeusz J\'{o}zefiak},
}

@incollection {mnev,
    AUTHOR = {Mn{\"e}v, N. E.},
     TITLE = {The universality theorems on the classification problem of
              configuration varieties and convex polytopes varieties},
 BOOKTITLE = {Topology and geometry---{R}ohlin {S}eminar},
    SERIES = {Lecture Notes in Math.},
    VOLUME = {1346},
     PAGES = {527--543},
 PUBLISHER = {Springer, Berlin},
      YEAR = {1988},
   MRCLASS = {52A25 (05B30 14G30 52A37)},
  MRNUMBER = {970093},
MRREVIEWER = {G. Ewald},
       DOI = {10.1007/BFb0082792},
       URL = {https://doi.org/10.1007/BFb0082792},
}

@article {BCT,
    AUTHOR = {Benedetti, Carolina and Chavez, Anastasia and Tamayo Jim\'{e}nez, Daniel},
     TITLE = {Quotients of uniform positroids},
   JOURNAL = {Electron. J. Combin.},
  FJOURNAL = {Electronic Journal of Combinatorics},
    VOLUME = {29},
      YEAR = {2022},
    NUMBER = {1},
     PAGES = {Paper No. 1.13, 20},
      ISSN = {1077-8926},
   MRCLASS = {05B35 (06A07)},
  MRNUMBER = {4395243},
MRREVIEWER = {Joseph\ E.\ Bonin},
       DOI = {10.37236/10056},
       URL = {https://doi.org/10.37236/10056},
    eprint = {1912.06873},
}

@article {asep,
    AUTHOR = {Corteel, Sylvie and Williams, Lauren K.},
     TITLE = {Tableaux combinatorics for the asymmetric exclusion process},
   JOURNAL = {Adv. in Appl. Math.},
  FJOURNAL = {Advances in Applied Mathematics},
    VOLUME = {39},
      YEAR = {2007},
    NUMBER = {3},
     PAGES = {293--310},
      ISSN = {0196-8858},
   MRCLASS = {05E10 (60C05)},
  MRNUMBER = {2352041},
MRREVIEWER = {Mikl\'{o}s B\'{o}na},
       DOI = {10.1016/j.aam.2006.08.002},
       URL = {https://doi.org/10.1016/j.aam.2006.08.002},
    eprint = {0602109}
}

@incollection {kps,
    AUTHOR = {Kodama, Yuji and Williams, Lauren},
     TITLE = {Combinatorics of {KP} solitons from the real {G}rassmannian},
 BOOKTITLE = {Algebras, quivers and representations},
    SERIES = {Abel Symp.},
    VOLUME = {8},
     PAGES = {155--193},
 PUBLISHER = {Springer, Heidelberg},
      YEAR = {2013},
   MRCLASS = {35C08 (14M15 35Q53)},
  MRNUMBER = {3183885},
       DOI = {10.1007/978-3-642-39485-0\_8},
       URL = {https://doi.org/10.1007/978-3-642-39485-0_8},
    eprint = {1205.1101},
}

@MastersThesis{Hmmath,
  author = {Simone Hu},
   title = {A combinatorial tale of two scattering amplitudes: see two bijections},
  school = {University of Waterloo},
    type = {MMath},
    year = {2021},
     DOI = {10012/17843},
  eprint = {2206.04749},
}

@article{agarwala_fryer_yeats_2022_I, 
    title={Combinatorics of the geometry of Wilson loop diagrams I: equivalence classes via matroids and polytopes}, 
    volume={74}, 
    DOI={10.4153/S0008414X21000134}, 
    number={4}, 
    journal={Canadian Journal of Mathematics}, 
    publisher={Canadian Mathematical Society}, 
    author={Agarwala, Susama and Fryer, Siân and Yeats, Karen}, 
    year={2022}, 
    pages={1177–1208},
    eprint={1908.10919},
}

@article{agarwala_fryer_yeats_2022_II, 
    title={Combinatorics of the geometry of Wilson loop diagrams II: Grassmann necklaces, dimensions, and denominators}, 
    volume={74}, 
    DOI={10.4153/S0008414X21000377}, 
    number={6}, 
    journal={Canadian Journal of Mathematics}, 
    publisher={Canadian Mathematical Society}, 
    author={Agarwala, Susama and Fryer, Siân and Yeats, Karen}, 
    year={2022}, 
    pages={1625–1672},
    eprint={1910.12158},
}
\medskip
\setlength\parskip{0pt}

\end{document}